\documentclass[aps,twocolumn,pra,superscript,floatfix,superscriptaddress,showpacs]{revtex4-1}

\usepackage{xcolor}
\usepackage{epsf}
\usepackage{epsfig}
\usepackage{graphicx}
\usepackage{dcolumn}
\usepackage{braket}
\usepackage{bm}
\usepackage{amsfonts}
\usepackage{amsmath}
\usepackage{amssymb}
\usepackage{color,soul}
\usepackage{wasysym}
\usepackage{mathrsfs}
\usepackage{natbib}
\usepackage{float}
\usepackage{multirow}

\usepackage[colorlinks=true,%
            linkcolor=blue,%
            urlcolor=blue,%
            citecolor=blue,%
            filecolor=blue,%
            bookmarksopen=true,%
            pdfauthor={J. P. Ramos-Andrade},%
            pdftitle={BIC},%
            pdfsubject={BIC_Manuscript},%
            pdfpagemode=UseOutlines]{hyperref}

\definecolor{otrogreen}{rgb}{0, 0.7, 0.}
\definecolor{antiquefuchsia}{rgb}{0.57, 0.36, 0.51}
\definecolor{amethyst}{rgb}{0.6, 0.4, 0.8}

\begin{document}

\title{Bound States in the Continuum in a Double Whispering Gallery Resonator}
\author{Alexis R. Leg\'{o}n}
\affiliation{Departamento de F\'{\i}sica, Universidad T\'{e}cnica Federico Santa Mar\'{i}a, Casilla 110 V, Valparaiso, Chile}
\author{M. Ahumada}
\affiliation{Departamento de F\'{\i}sica, Universidad T\'{e}cnica Federico Santa Mar\'{i}a, Casilla 110 V, Valparaiso, Chile}
\affiliation{Departamento de F\'isica, Universidad de Santiago de Chile, Avenida V\'ictor Jara 3493, 9170124, Santiago, Chile}
\author{J. P. Ramos-Andrade}
\affiliation{Departamento de F\'isica, Universidad de Antofagasta, Av. Angamos 601, Casilla 170, Antofagasta, Chile.}
\author{Rafael A. Molina}
\affiliation{Instituto de Estructura de la Materia – CSIC, Serrano, 123. E-28006 Madrid, SPAIN}
\author{P.\ A.\ Orellana}
\affiliation{Departamento de F\'{\i}sica, Universidad T\'{e}cnica Federico Santa Mar\'{i}a, Casilla 110 V, Valparaiso, Chile}
\

\begin{abstract}
In this work, we investigate the single-photon transport through two whispering gallery resonators (WGRs) coupled to a one-dimensional waveguide. Using Green's function formalism, we compute the transmission spectra and the photonic density of states (DOS) for the stationary states. We also obtain the formation of two types of bound states in the continuum (BICs). 
The first kind is localized  into the  WGR and are symmetry-protected BICs. In contrast, the second depends on the distance between resonators through the waveguide and is of the Fabry-Perot kind.
These BICs are represented as Dirac delta functions in the local density of states. Additionally, we show that quasi-BICs manifest as sharp resonances in photonic transmission due to small symmetry-breaking perturbations.
Furthermore, we investigate the dynamics of a single-photon wave packet interacting with the WGRs and analyze the mechanism for storing the wave packet in the structure formed by the WGRs and the finite  waveguide between them.
\end{abstract}

\maketitle

\section{Introduction}

A bound state in the continuum (BIC) 
coexists with extended waves in the continuum,
yet remains perfectly confined without any radiation leakage. BICs are found in a wide range of material systems through confinement mechanisms that are fundamentally different from those of conventional bound states \cite{Hsu2016}. In 1929, BICs were proposed by von Neumann and Wigner in a quantum system with a confining oscillatory potential \cite{vonNeuman1929}. 
More recently,
a range of alternative mechanisms for achieving BICs have been uncovered in various material systems. Many of these mechanisms have been experimentally observed in various contexts, demonstrating their versatility and broad applicability across different types of material systems \cite{Hsu2016, Joseph2021, Zhang2023}. In recent years, photonic structures have become an especially appealing platform because they allow for customization of the material and structure, a capability often not feasible in quantum systems \cite{Sadriev2019, Chai2024}. The distinctive properties of BICs have resulted in numerous applications, including lasers, sensors, filters, and low-loss fibers, with many additional potential uses proposed and yet to be realized \cite{Kang2021, Joseph2021, Khattou2023}.

Resonators are crucial in these advancements involving BICs. 
Resonators, designed to sustain or amplify oscillations, are vital in many scientific and engineering fields.
They come in various types, each suited for specific applications \cite{Ossiander2023}. Fabry-Pérot resonators use parallel mirrors to create standing waves commonly found in laser cavities. Ring resonators utilize circular waveguides for continuous wave looping, making them useful in optical communications \cite{Xu2007, Afroozeh2018, Kazakov2024}. Photonic crystal resonators employ a periodic structure to confine light through Bragg scattering \cite{Rybin2010}, achieving high-quality resonances \cite{Blanchard2016, Athe2016}. Among these, whispering gallery resonators (WGRs) are unique for their ability to confine waves along the curved surface of a circular or spherical structure via total internal reflection \cite{Vahala2003, Koshelev2020}. This allows WGRs to achieve exceptionally high-quality factors \cite{Yan2015, Yoshiki2017, Perez2023, Suebka2024}, making them ideal for sensing \cite{Armani2003}, filtering \cite{Kippenberg2004}, and nonlinear optics applications \cite{Lin2014}.


Exploring new contexts for observing BICs is a key area of interest. Single-photon transport provides an excellent platform for such investigations. Additionally, photons offer advantages over other information carriers, such as electrons, by enabling high-speed transmission and exhibiting low dissipation in nonlinear media \cite{Qin2016,Ahumada2018,Feng2021}. However, when considering emitters separated by large distances, retardation effects due to the finite speed of light can play a significant role, as discussed in the study of collective radiation from distant emitters \cite{Tudela2020}. These effects can lead to phenomena such as linewidth broadening and Fano-resonance-like peaks, further enriching the study of BICs in photonic systems \cite{Tudela2020}. Typically, an ideal WGR displays a symmetric spectrum due to the degeneracy between the rotating and counter-rotating modes. This inherent symmetry can facilitate the formation of BICs within WGRs when they are coupled to a continuum \cite{Ahumada2018}. A symmetry-breaking element needs to be introduced into the system to enable the coupling of the BIC to the continuum. They become the quasi-BICs, typically acquiring a Fano resonance shape controllable by the perturbation \cite{Guevara2003}.

In this study, we explore single-photon transport through two WGRs coupled to a one-dimensional waveguide, uncovering the emergence of BICs between the WGRs. We discovered that when the separation distance $L$ between the WGRs is even, BICs manifest as sharp resonances in the photonic transmission. We also obtained BICs for even $L$ by breaking the frequency symmetry. Furthermore, we investigated the time evolution of an incident single-photon wave packet interacting with the WGRs. We analyzed the mechanism of partial storage of the wave packet in the BIC. These findings highlight the potential of WGRs and BICs in enhancing single-photon transport and storage in photonic systems.



\section{Model}
\label{sec:1}


\begin{figure}[h]
\includegraphics[width=\columnwidth]{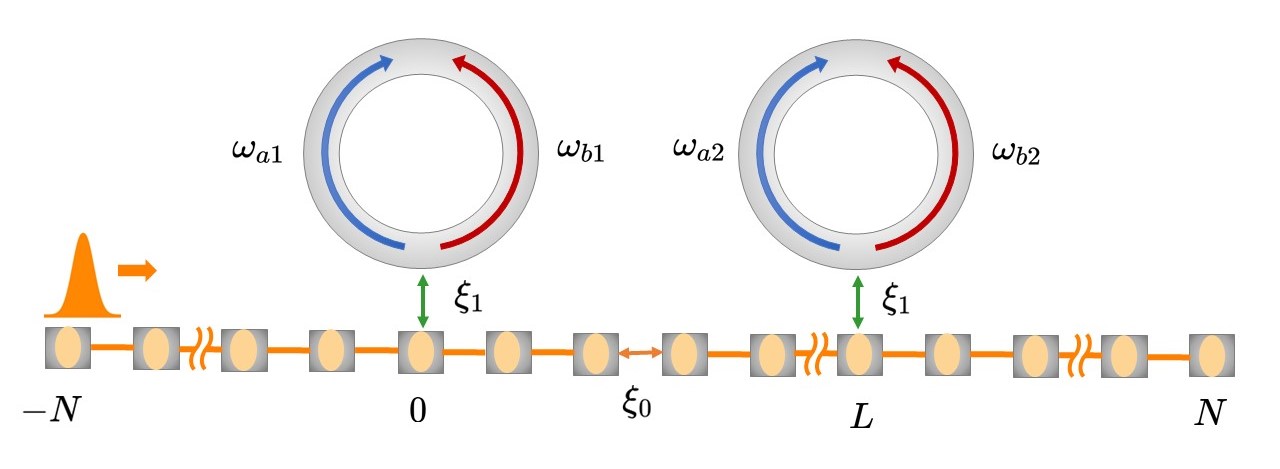}
\caption{Schematic representation of the system: A one-dimensional waveguide functioning as a transmission line, which is coupled to two resonators. The solid blue arrows (\(\omega_{a1}\), \(\omega_{a2}\)) 
and the solid red arrows (\(\omega_{b1}\), \(\omega_{b2}\)) 
represent the clockwise and counter-clockwise modes associated with the first and second resonators, respectively. The green arrows indicate the coupling \(\xi_{1}\) between the waveguide and the resonators. The orange line depicts the hopping \(\xi_{0}\) between two adjacent cavities in the discrete waveguide.}
\label{fig1}
\end{figure}

A scheme of the system of two WGRs coupled through a waveguide modeled by a one-dimensional tight-binding lattice is represented in Fig. \ref{fig1}. The Hamiltonian of the system can be written as $H=H_{0}+H_{\text{I}}$, where $H_{0}=H_{\text{W}}+H_{\text{WG}}$ corresponds to the sum of the contributions of the waveguide $H_{\text{W}}$ and the WGR modes $H_{\text{WG}}$, respectively. The explicit expressions of $H_{\text{W}}$ and $H_{\text{WG}}$ are 
\begin{eqnarray}
H_{\text{W}}&=&\omega_{c}\sum_{j=-\infty}^{\infty}\hat{c}_{j}^{\dag}\hat{c}_{j}-\xi_{0}\sum_{j=-\infty}^{\infty}(\hat{c}_{j}^{\dag}\hat{c}_{j+1}+\hat{c}_{j+1}^{\dag}\hat{c}_{j}) \, ,
\label{1.a} \\  
H_{\text{WG}}&=& \sum_{i=1,2}(\omega_{ai}\,\hat{a}_{i}^{\dag}\hat{a}_{i}+\omega_{bi}\,\hat{b}_{i}^{\dag}\hat{b}_{i})\,, 
\label{1.b} 
\end{eqnarray}
and $H_{\text{I}}$ represents the interactions between the waveguide and the WGR modes. The expression for $H_{\text{I}}$ is
\begin{eqnarray}
H_{\text{I}}=&&- \xi_{1}(\hat{c}_{0}^{\dag}\hat{a}_{1}+\hat{c}_{0}\hat{a}_{1}^{\dag}+\hat{c}_{0}^{\dag}\hat{b}_{1}+\hat{c}_{0}\hat{b_{1}}^{\dag})\nonumber \\ 
&&-\xi_{1}(\hat{c}_{L}^{\dag}\hat{a}_{2}+\hat{c}_{L}\hat{a}_{2}^{\dag}+\hat{c}_{L}^{\dag}\hat{b}_{2}+\hat{c}_{L}\hat{b}_{2}^{\dag})\,,  
\label{1.c} 
\end{eqnarray}
where $\hat{c}_{j}^{\dag}$ ($\hat{c}_{j}$) are the creation (annihilation) operator of a single photon of the $j$-th waveguide site with frequency $\omega_{c}$, and
$\xi_{0}$ is the hopping coefficient between any two nearest-neighbor sites in the waveguide. The operators
$\hat{a}^{\dag}$ ($\hat{a}$) and $\hat{b}^{\dag}$ ($\hat{b}$) represent the creation (annihilation) operators for clockwise (CW) and counter-clockwise (CCW) resonator modes with frequencies $\omega_{a}$ and $\omega_{b}$, respectively. The coupling between the zeroth and $L$-th waveguide sites 
and both resonators is $\xi_{1}$.
The dispersion relation of the 1-D discrete waveguide is given by $\omega=\omega_{c}-2\xi_{0}\cos(k)$ with $k \in [0,2\pi]$, resulting in energy bandwidth $4\xi_{0}$.

\section{Analyzing Transmission Spectra and Bound States}
\label{sec:2}
We now explore our system's behavior in stationary state, determining the transmission probability and density of states (DOS). We utilize the Lippmann-Schwinger and Dyson equations \cite{Economou2006, Mahan2000} to compute Green's function and, consequently, the transmission amplitude and DOS of the system. \ref{App.A}.

\


Following the procedure outlined in Appendix \ref{App.A}, we derive the expression for the transmission amplitude as follows
\begin{equation}
t=\frac{1}{1-(g_{w_1}+g_{w_2})\Sigma_{0} +g_{w_1}g_{w_2}\Sigma_{0}^{2}(1-e^{2ikL})} \, ,
\label{2.8}
\vspace{1mm}
\end{equation}
where $g_{w_1}=1/(\omega-\omega_{a1}+i\eta) +1/(\omega-\omega_{b1}+i\eta)$ and $g_{w_2}=1/(\omega-\omega_{a2}+i\eta) +1/(\omega-\omega_{b2}+i\eta)$, with $\eta \rightarrow 0$, correspond to the Green's functions of the WGRs and $\Sigma_{0}=\xi_{1}^{2}g_{0}=\xi_{1}^{2}/2i\xi_{0}\sin(k)$ to the self-energy.

The transmission probability, defined as $T =t \cdot t^{*} =|t|^{2}$, is given by
\begin{widetext}
\begin{equation}
T=\frac{4 \xi_{0}^{4}\sin^{4}(k)}{\left[g_{w_1} g_{w_2}\sin^{2}(kL)-2\xi_{0}^{2}\sin^{2}(k)\right]^{2} + \left[\xi_{0}(g_{w_1}+g_{w_2})\sin(k)-g_{w_1}g_{w_2}\sin(kL)\cos(kL)\right]^{2}} \, .
\label{2.9}
\end{equation}
\end{widetext}

The above Eq.\,(\ref{2.9}) illustrates the complex interactions between wave propagation and device architecture, essential for understanding transmission efficiency and resonance in photonic systems. This calculate transmission probability through the system, highlighting the interplay between the wave vector \(k\), Green's functions \(g_{w_1}\) and \(g_{w_2}\), and intrinsic properties \(\xi_{0}\). This analysis reveals how these factors collectively influence the system's transmission efficiency based on underlying physical principles.

One of the main advantages of using Green's function is the straightforward methodology for calculating the density of states through the imaginary part ($\Im$) of the trace of the Green's function. We focus on the local density of states (LDOS), where we trace only over the sites of interest.
The expression is, then, as follows:
\begin{equation}
\rho_{i}(\omega) = -\frac{1}{\pi} \, \Im \left[  G_{i,i} (\omega) \right] \, .
\label{2.10}
\end{equation}

Accordingly, the total DOS can be summed up from the LDOS contributions across 
the whole system.
represented by:
\begin{equation}
\rho_{T}(\omega) = \sum\limits_{i} \, \rho_{i} (\omega) \, , 
\label{2.11}
\end{equation}
This method streamlines the process of dissecting the system's spectral features.

Revisiting our system of two WGRs, and bearing the previously mentioned expressions in mind, the sole requirement to ascertain the DOS associated with the WGRs and between them is to compute the trace of the Green's function 
for the relevant location.
This calculation is crucial for identifying the presence or emergence of BICs, which are localized states that remain trapped and do not couple to the continuum despite the overlapping spectral range. 


Now, to translate this understanding into a quantitative framework, let us consider the formula for calculating the LDOS in the resonators for our system of coupled WGRs. Using the corresponding Green's functions of our coupled system and the expressions in Eqs. (\ref{2.10}) and (\ref{2.11}), we obtain the DOS:
\begin{equation}
\rho_{w_1 +w_2}(\omega) =  -\frac{1}{\pi} \, \sum\limits_{l,m} \, \Im \left[  G_{l,l} (\omega) + G_{m,m} (\omega) \right] \, ,
\label{2.12}
\end{equation}
the summation is carried out over the modes of the WGRs, specifically $l=\omega_{a1},\omega_{b1}$ for the first WGR, and $m=\omega_{a2},\omega_{b2}$ for the second WGR, respectively. Thus, $\rho_{w_1+w_2}(\omega)$ is calculated using the inverse of the Green's function for the WGRs, as detailed in Appendix \ref{App.B}:
\begin{subequations}
\begin{eqnarray}
G_{l,l}(\omega)&= \dfrac{g_{l}\, (1-  g_{w_2} \, g_{0} \, \xi_{1}^{2})}{1-(g_{l}+g_{w_2})\Sigma_{0} +g_{l}\,   g_{w_2} \Sigma_{0}^{2}(1-e^{2ikL})}  \, ,\nonumber \,\\ \label{2.13.a} \\
G_{m,m}(\omega)& = \dfrac{g_{m}\, (1-  g_{w_1} \, g_{0} \, \xi_{1}^{2})}{1-(g_{m}+g_{w_1})\Sigma_{0} +g_{m}\,   g_{w_1} \Sigma_{0}^{2}(1-e^{2ikL})} \, , \nonumber \, \\
\label{2.13.b}
\end{eqnarray}
\end{subequations}
where $g_{m} = 1/(\omega - m + i \eta)$ and $g_{l} = 1/(\omega - l + i \eta)$. 
This methodical approach guarantees a thorough analysis of the resonant frequencies specific to each WGR, providing the necessary spectral insights to understand resonance behavior and the conditions conducive to the emergence of BIC within the system.

To assess the potential locations of BIC between the WGRs, we define the corresponding DOS as 
\begin{equation}
\rho_{{w_1}\leftrightarrow{w_2}}(\omega) =  -\frac{1}{\pi} \, \sum\limits_{j=0}^{L} \, \Im \left[  G_{j,j} (\omega) \right] \, ,
\label{2.14}
\end{equation}
 which involves the use of the specific Green's function between the WGRs, detailed in Appendix \ref{App.C}: 
\begin{equation}
G_{j,j}= \frac{ g_{0}  \left( 1 -F_{j} \right) }{1-(g_{w_1}+g_{w_2})\Sigma_{0} +g_{w_1}\,   g_{w_2} \Sigma_{0}^{2}(1-e^{2ikL})} \, ,
\label{2.15}
\end{equation}
where $F_{j}$ is defined as:
\begin{eqnarray}
F_{j} & \equiv g_{w_1} \, \Sigma_{0} \, (1-e^{2ikj}) +g_{w_2} \, \Sigma_{0} \, (1-e^{2ik(L-j)}) \nonumber \\
&- g_{w_1} \, g_{w_2} \,  \Sigma_{0}^{2} (1+e^{2ikL}) (1+e^{-2ikj}) \, .
\label{2.16}
\end{eqnarray}
This formulation provides a methodological framework for pinpointing the exact conditions under which BICs emerge between WGRs, by analyzing the LDOS through the cumulative effect of interactions across the waveguide, where the WGRs are spaced apart by a distance of length $L$ (see Fig.\,\ref{fig1}). 

\section{Time-Dependent Photon Dynamics}
\label{sec:3}
We study the temporal evolution of the system by solving the time-dependent Schr\"odinger equation for the Hamiltonian $H$ for a given initial condition. In particular, we analyze the BIC when coupled or uncoupled to the continuum states and the effects that this coupling manifests in the temporal evolution of the initial condition. 

We initiate our analysis with a Gaussian photon wave packet impinging on the system, described by:
\begin{equation}
\psi(x,0)=\frac{1}{\sqrt{\sigma\sqrt{\pi}}}\;e^{-{\frac{(x-x_{0})}{2\sigma^{2}}^{2}}}e^{ik_{0}x} \, ,
\label{3.1}
\end{equation}
 where $\sigma$, $k_{0}>0$, and $x_{0}$ are the width, the wave vector, and the initial position of the center of the wave packet, respectively. We  consider the incident pulse propagating from left to right in the input channel, which can be partially reflected back and partially transmitted into the output channel (see Fig.\,\ref{fig1}). 

 Another initial condition (at $t=0$), we evaluate for our system is within the WGRs. For the first WGR, we consider the anti-symmetry condition of modes,
\begin{equation}
|\Psi_{w_1}\rangle= \frac{1}{\sqrt{2}}\, ( \hat{a}_{1}^{\dag}-\hat{b}_{1}^{\dag} ) |0\rangle \, . 
\label{3.2}
\end{equation}

Taking an anti-symmetric state results in the localization of modes at the extremes of the WGRs. Specifically,
\begin{equation}
|\Psi_{w_{1}\leftrightarrow{w_2}}\rangle= \frac{1}{\sqrt{2}}\, ( \hat{a}_{1}^{\dag}-\hat{b}_{2}^{\dag} ) |0\rangle \, , 
\label{3.3}
\end{equation}
where it corresponds to the first mode of the first WGR and the second mode of the second WGR. 

For Eqs.\,(\ref{3.2}) and (\ref{3.3}), we consider breaking symmetry  in the Hamiltonian $H_{\text{WG}}$ of Eq.\,(\ref{1.b}). For this, we introduce a term $\Delta$ in the frequencies of WGR modes  (with $\Delta \ll \omega_{ai} \, , \omega_{bi})$ in two forms:
\begin{subequations}
\begin{eqnarray}
H^{\text{Intra}}_{\text{WG}}&=& (\omega_{a1} +\Delta)\,\hat{a}_{1}^{\dag}\hat{a}_{1}+(\omega_{b1}-\Delta)\,\hat{b}_{1}^{\dag}\hat{b}_{1} \nonumber \\
&&+(\omega_{a2}+\Delta)\,\hat{a}_{2}^{\dag}\hat{a}_{2}+(\omega_{b2}-\Delta)\,\hat{b}_{2}^{\dag}\hat{b}_{2}\,, \label{3.4.a} \\
H^{\text{Inter}}_{\text{WG}}&=& (\omega_{a1} +\Delta)\,\hat{a}_{1}^{\dag}\hat{a}_{1}+(\omega_{b1}+\Delta)\,\hat{b}_{1}^{\dag}\hat{b}_{1} \nonumber \\
&&+(\omega_{a2}-\Delta)\,\hat{a}_{2}^{\dag}\hat{a}_{2}+(\omega_{b2}-\Delta)\,\hat{b}_{2}^{\dag}\hat{b}_{2}\, .\label{3.4.b} 
\end{eqnarray}
\end{subequations}
When applying the symmetry-breaking schemes  described in Eqs.\,(\ref{3.4.a}) and (\ref{3.4.b}), BICs emerge.We analyze the dynamics of the system numerically using the TKWANT Python library \cite{Kloss_2021}.
Our analysis of temporal evolution was structured into two main parts: the examination of Gaussian wave packets, and the exploration of anti-symmetry states $|\Psi_{w_1}\rangle$ and $|\Psi_{w_{1}\leftrightarrow{w_2}}\rangle$ when considering the symmetry-breaking of WGRs [see Eqs.\,(\ref{3.4.a}) and (\ref{3.4.b})], which facilitates communication between them. It is important to note that our analysis focused specifically on cases where the separation distance $L$ is even. This arrangement is crucial, as it is under these precise circumstances that BICs manifest, a phenomenon corroborated by our analysis of the stationary state.


To investigate the probability transfer in the system dynamics, we define \textit{the transfer fidelity} $(TF)$ between the WGRs as follows:
\begin{equation}
\begin{small}
TF = \frac{1}{2} \left( \left| \langle \psi(t) | a_i^\dagger a_j | \psi(0) \rangle \right|^2 + \left| \langle \psi(t) | b_i^\dagger b_j | \psi(0) \rangle \right|^2 \right) \, ,
\label{3.5}
\end{small}
\end{equation}
where \(\psi(t)\) and \(\psi(0)\) represent the state of the system at times \(t\) and \(0\), respectively. The operators \(a_i^\dagger a_j\) and \(b_i^\dagger b_j\) correspond to the creation and annihilation operators for modes \(i\) and \(j\) within the WGRs. Specifically, \(a_i^\dagger a_j\) acts on the mode \(j\) to transfer an excitation to mode \(i\) within one WGR, while \(b_i^\dagger b_j\) does the same within another WGR. This expression allows us to quantify the fidelity of probability transfer, capturing the interactions between different modes within the WGRs.

\section{Results}
\label{sec:results}

\begin{figure}[htp]
\includegraphics[width=\columnwidth]{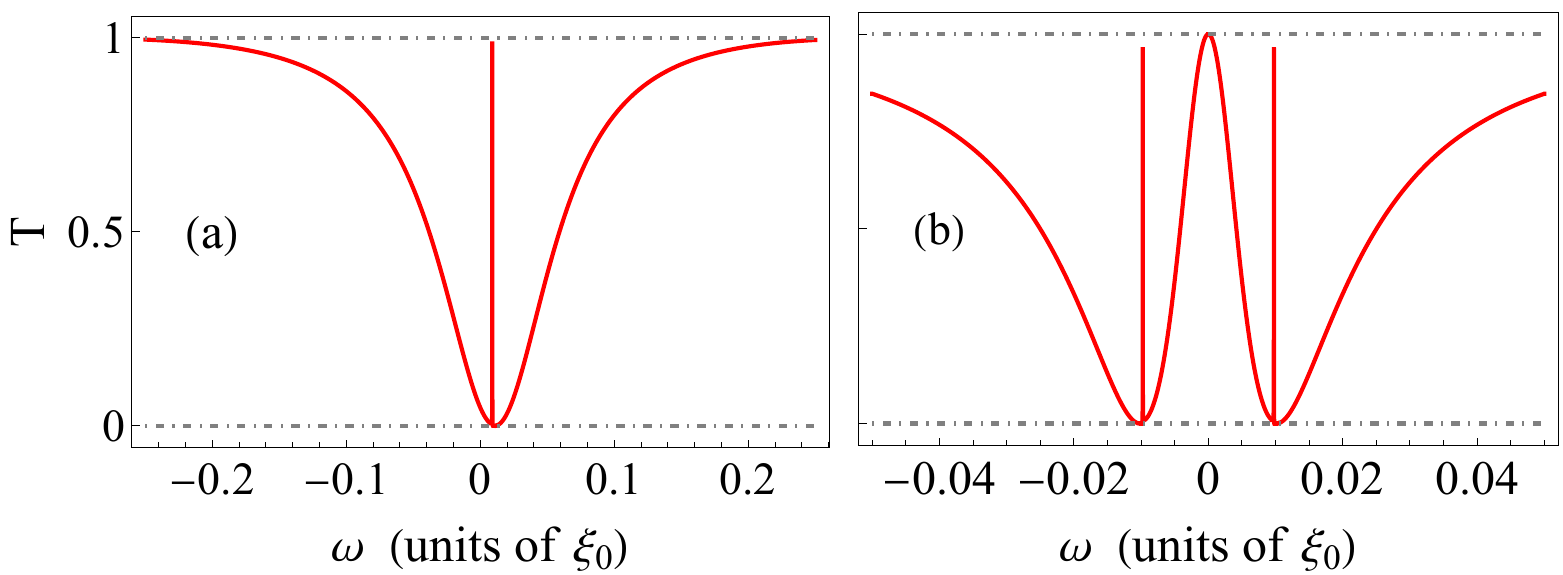}
    \caption{Transmission probability as function of $\omega$. (a) $\omega_{a1}=\omega_{b1}= \omega_{a2}=\omega_{b2}=0.01$ and (b) $\omega_{a1}=-\omega_{b1}= \omega_{a2}=-\omega_{b2}=0.01$ are resonance frequencies. The parameters are chosen as $\xi_1=0.1$  and $\xi_0=1$ (all in units of $\xi_0$ ) and $\omega_{c}=0$. }
    \label{fig2}
\end{figure}

In the stationary state regime, we show the transmission probability in Fig.\,\ref{fig2}, demonstrating characteristic Fano symmetric antiresonance behavior in photon propagation through such systems. This Fano symmetric line shape is disturbed by a sharp resonance due to the formation of quasi-BICs for even $L$ values, as seen in Figs.\,\ref{fig2}(a) and \ref{fig2}(b) for origin and anti-resonances respectively.
The emergence of BICs significantly boosts photon transmission by interacting with the continuous spectrum.
These states emerge from the interference among degenerate states within the WGRs and the continuous states in the waveguide, reminiscent of the Fano effect \cite{Fano1866,Miroshnichenko2010}.

\begin{figure}[hbp]
\includegraphics[width=\columnwidth]{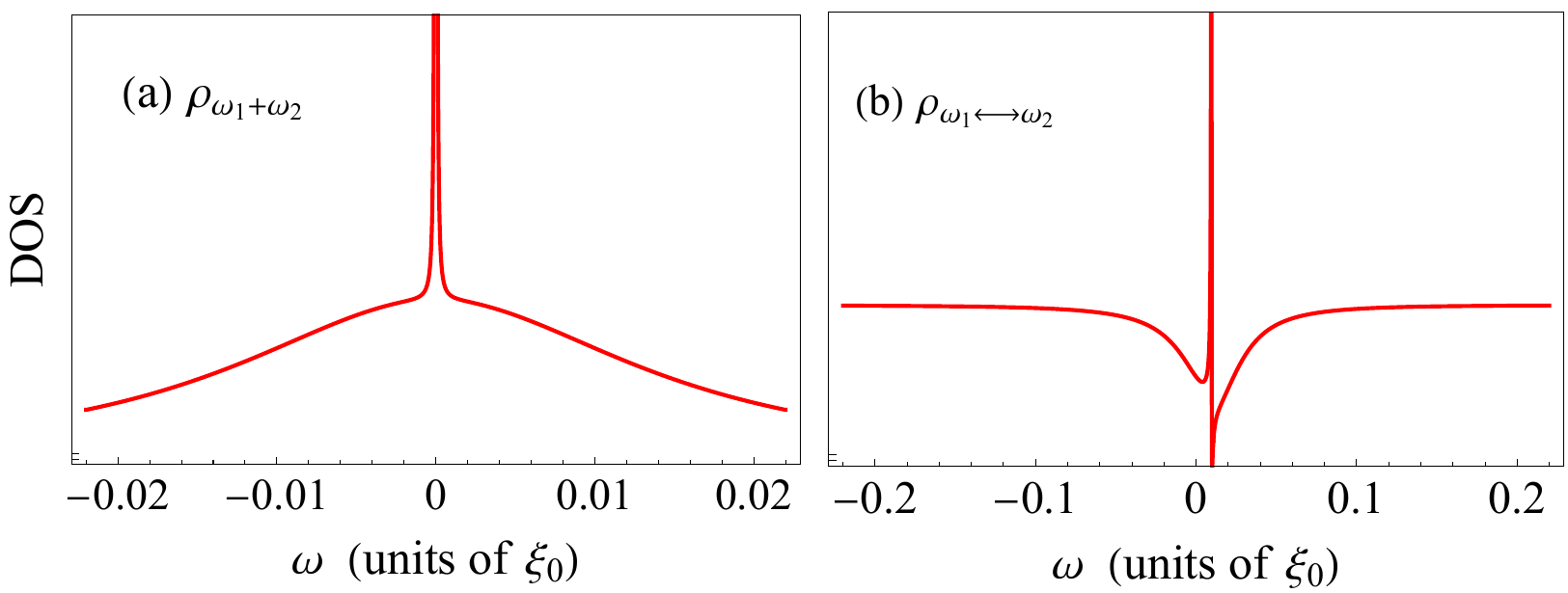}
    \caption{(a) DOS $\rho_{w_1+w_2}$ and (b) DOS $\rho_{w_1 \leftrightarrow w_2}$ as function of $\omega$. The parameters are chosen as $\xi_1=0.1$  and $\xi_0=1$ (all in units of $\xi_0$ ), $\omega_{c}=0$, $\omega_{a1}=\omega_{b1}= \omega_{a2}=\omega_{b2}=0.01$  are resonance frequencies.}
    \label{fig3}
\end{figure}

In the DOS analysis for WGRs, using the expressions $\rho_{w_1+w_2}(\omega)$ and $\rho_{w_1 \leftrightarrow w_2}(\omega)$, we see that the usual symmetry in the system is interrupted by the emergence of BICs, as highlighted in Fig.\,\ref{fig3}(a). The presence of BICs introduces Dirac $\delta$-like peaks in the spectral analysis, indicating highly localized, uncoupled modes within the continuum. This modification, evident in the figures referenced, drastically affects the system's spectral properties and alters mode interactions—critical for applications demanding precise energy and resonance control. 



Additionally, within the interaction DOS $\rho_{w_1 \leftrightarrow w_2}(\omega)$ for the WGRs [Fig.\,\ref{fig3}(b)], the typical symmetry associated with the formation of a Fano resonance for the BIC between the WGRs. This alteration points to a pivotal interaction within our setup. Moreover, this Fano resonance introduces a sharp, delta-like peak in the spectrum, serving as a clear marker for identifying the presence of BICs. Such interactions and the resulting sharp resonance phenomena are critical for understanding the intricate dynamics that govern the behavior of photonic systems involving WGRs, highlighting the delicate balance between coupled and uncoupled modes that characterize these systems.


\begin{figure}[h]
\includegraphics[width=\columnwidth]{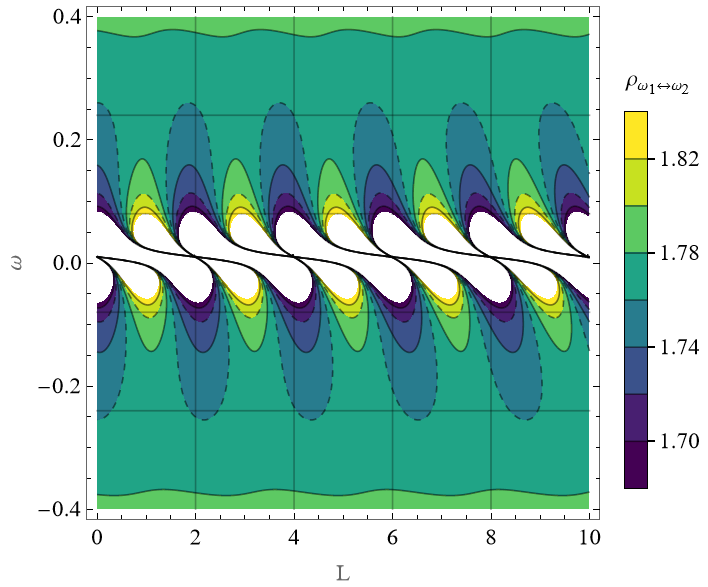}
\caption{DOS among WGRs, as a function of frequency $\omega$  and the site number $L$. The parameters are chosen as $\xi_1=0.1$  and $\xi_0=1$ (all in units of $\xi_0$ ), $\omega_{c}=0$, $\omega_{a1}=\omega_{b1}= \omega_{a2}=\omega_{b2}=0.01$  are resonance frequencies. }
\label{fig4}
\end{figure}

The analysis of the DOS as a function of frequency $\omega$ and the site number between the WGRs ($L$), as depicted in Fig.\,\ref{fig4}, uncovers distinct sharp peaks at even $L$ values, signaling the presence of BICs. These peaks, accentuated by the intensity of the colors on the scale, represent light trapped within resonant modes due to constructive interference, leading to highly localized states that, surprisingly, do not radiate energy. This not only indicates a higher DOS but also underscores the inherent symmetry of the system that fosters the formation of BICs. Such symmetry, crucial in the waveguide system, establishes a resonance pattern at even \text{$L$} intervals, which is vital for photonic applications requiring precise control over light propagation.

The observed BICs, resulting from the constructive interference of degenerate states at identical energy levels, are especially pronounced for pairs of WGRs at even numbers of $L$, reflecting a specific resonant or symmetrical arrangement. These dynamics between WGR pairs are critical for designing photonic devices capable of precise light manipulation. Additionally, the white areas visible in the graph correspond to regions that exceed the color scale range, signifying extremely high DOS values. These zones highlight the surroundings of the Fano resonance where the BICs are located, specifically at the intersections of the lines for even values of \text{$L$}. This detailed observation underscores the critical interplay between WGR pairs in photonic device design, essential for achieving meticulous light manipulation in photonic applications.



\begin{figure}[h]
\includegraphics[width=\columnwidth]{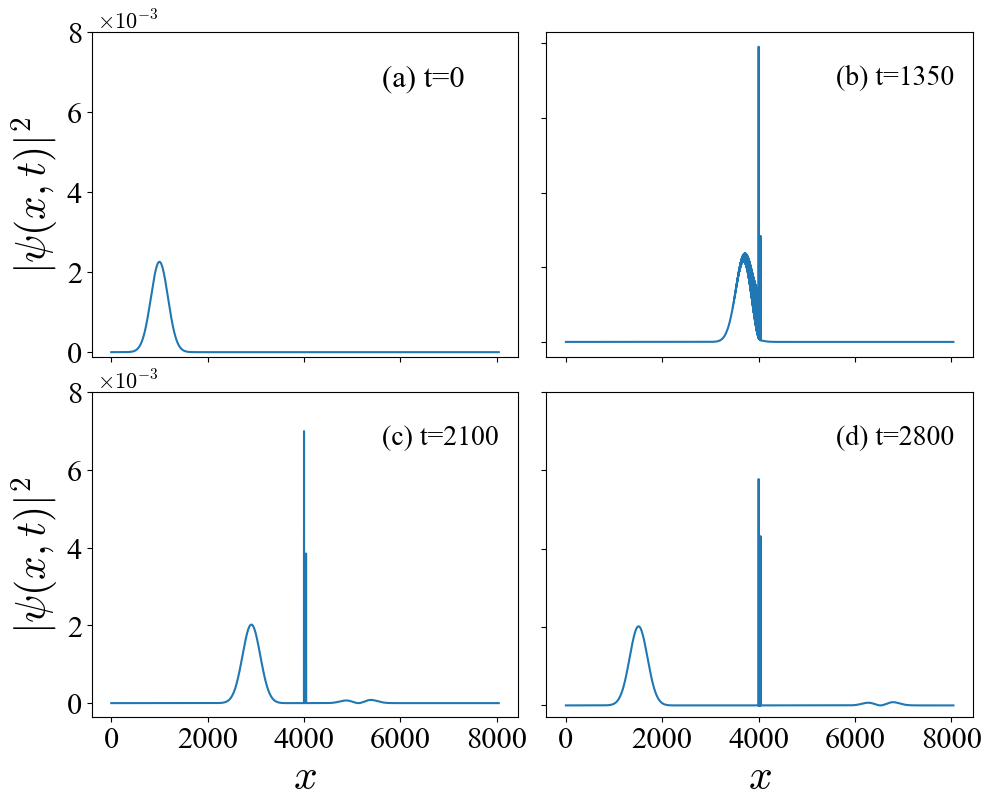}
    \caption{Probability density of the wave packet as a function of the site $x$ of the waveguide. The temporal evolution is shown for (a) $t=0$, (b) $t=1350$, (c) $t=2100$ and (c) $t=2800$ ($t$ in units of $\hbar / \xi_{0}$). The parameters are chosen as $\xi_1=0.1$ and $\xi_0=1$ (all in units of $\xi_0$), $\omega_{c}=0$, $\omega_{a1}=-\omega_{b1}= \omega_{a2}=-\omega_{b2}=0.001$ are the resonance frequencies and $L=40$ and $N=4000$.}
    \label{fig5}
\end{figure}

We observe the evolution of anti-symmetric frequencies for the Gaussian wave packet depicted in Fig.~\ref{fig5}. The initial conditions are defined by $\sigma=250\, ,x_0=4000\, ,k_0=\pi/2$, and $N=4000$, where $N$ represents the number of sites coupled to the sides of the WGRs. Figure~\ref{fig5}(a) illustrates the initial Gaussian wave packet. Subsequently, the pulse propagates through the system and scatters off the WGRs, as depicted in Figs.~\ref{fig5}(b) and \ref{fig5}(c). In Fig.~\ref{fig5}(d), the probability density reaches its maximum value, indicating the occupation of the WGRs. Furthermore, Fig.~\ref{fig5}(d) demonstrates that some probability remains trapped in the WGRs while the remainder of the pulse begins to spread along the waveguide.

\begin{figure}[htp]
\includegraphics[width=\columnwidth]{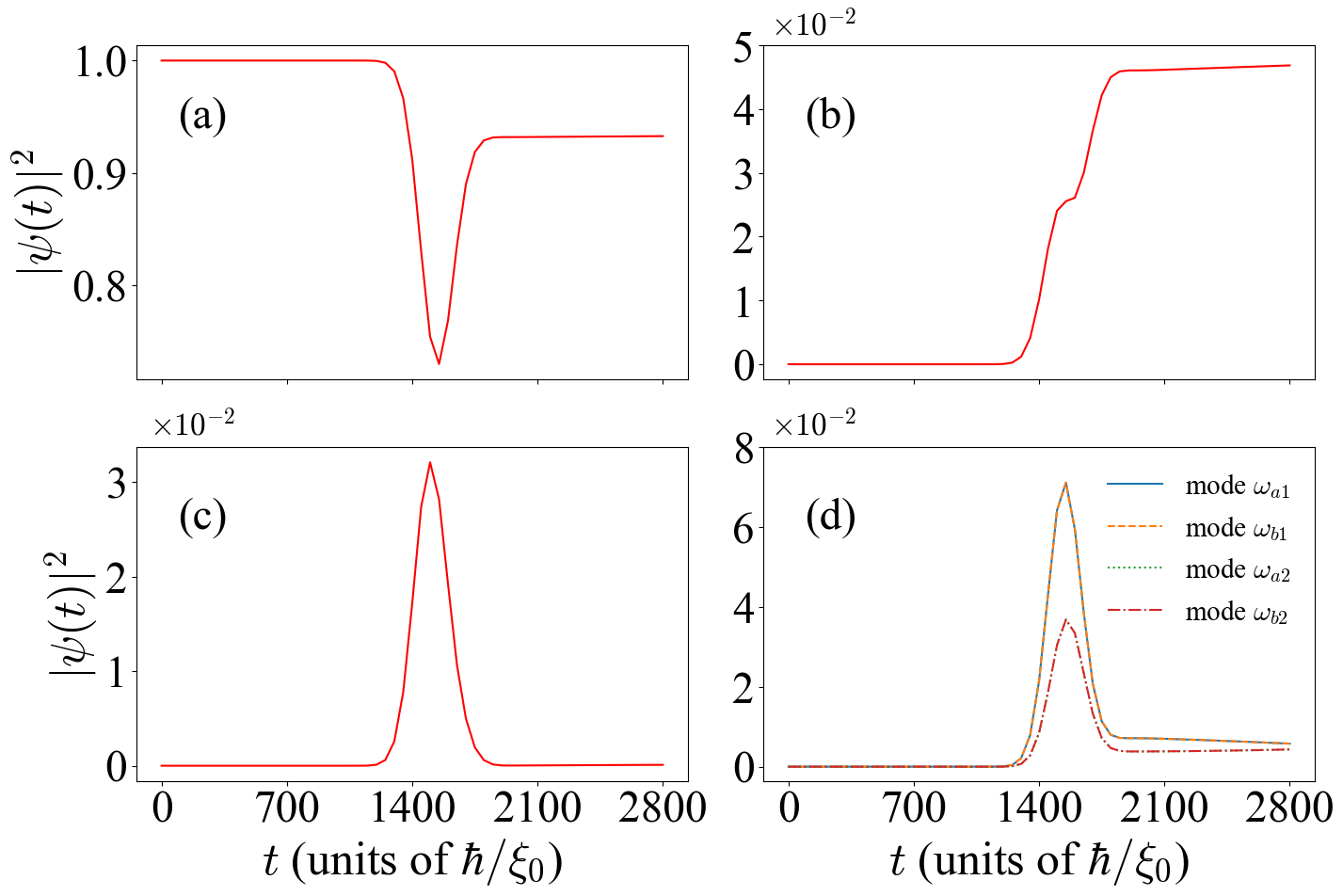}
     \caption{Probability density of the wave packet as a function of the time $t$ (in units of $\hbar / \xi_{0}$). The probability density for: (a) the left side of the WGRs; (b) the right side of the WGRs; (c) between the WGRs, and; (d) for the different modes of the WGRs. The parameters are chosen as $\xi_1=0.1$ and $\xi_0=1$ (all in units of $\xi_0$), $\omega_{c}=0$, $\omega_{a1}=-\omega_{b1}= \omega_{a2}=-\omega_{b2}=0.001$ are the resonance frequencies and $L=40$ and $N=4000$.}
    \label{fig6}
\end{figure}

Upon inspecting the mode occupancy in the WGRs, as illustrated in Fig.~\ref{fig6}(d), a significant increase in occupancy in the modes of the first WGR is observed following the passage of the Gaussian wave packet through the WGRs. Similarly, as depicted in Fig.~\ref{fig6}(c), a notable spike in probability density is observed during the interval in which the wave packet traverses the system, subsequently diminishing. The remaining probability density is dispersed among the modes of the WGRs as well as across the regions to the left [Fig.~\ref{fig6}(a)] and right [Fig.~\ref{fig6}(b)] of the WGRs.

\begin{figure*}[htbp]
\includegraphics[width=\textwidth]{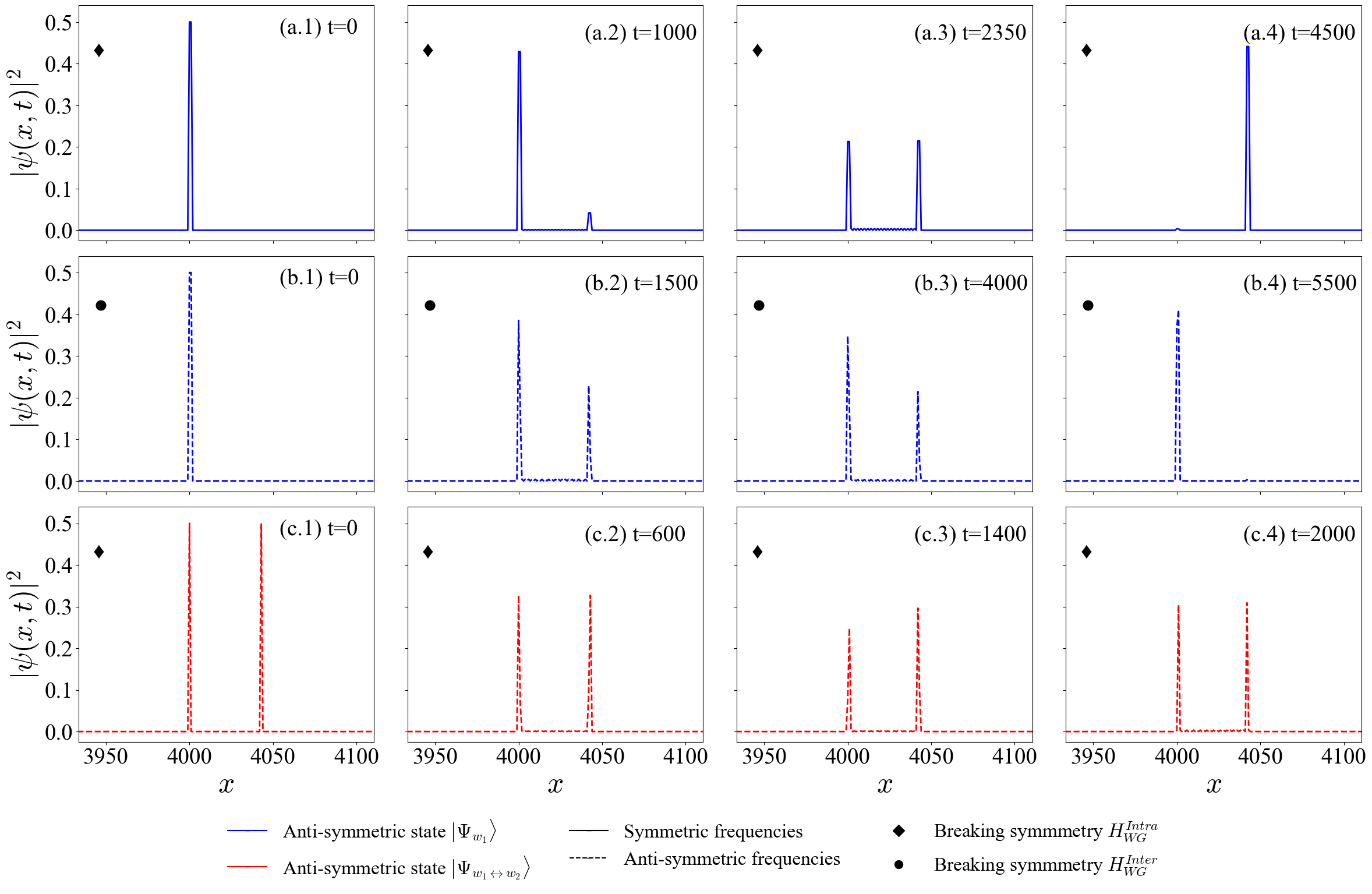}
    \caption{Probability density from the initial conditions for the anti-symmetric states (\(|\Psi_{w_1}\rangle\) and \(|\Psi_{w_{1}\leftrightarrow{w_2}}\rangle\)) as a function of the site \(x\) in the waveguide (Breaking symmetry with $H^{\text{Intra}}_{\text{WG}}$ and $H^{\text{Inter}}_{\text{WG}}$ ). The temporal progression is displayed as follows: in the first row, (a.1) $t=0$, (a.2) $t=1000$, (a.3) $t=2350$, and (a.4) $t=4500$ (units of $\hbar / \xi_{0}$); the second row illustrates (b.1) $t=0$, (b.2) $t=1500$, (b.3) $t=4000$, and (b.4) $t=5500$ (units of $\hbar / \xi_{0}$); finally, the third row showcases (c.1) $t=0$, (c.2) $t=600$, (c.3) $t=1400$, and (c.4) $t=2000$ (units of $\hbar / \xi_{0}$). The parameters are set as $\xi_1=0.1$ and $\xi_0=1$ (both in units of $\xi_0$), with $\omega_{c}=0$. For symmetric frequencies, $\omega_{a1}=\omega_{b1}= \omega_{a2}=\omega_{b2}=0.001$, while for anti-symmetric frequencies, $\omega_{a1}=-\omega_{b1}= \omega_{a2}=-\omega_{b2}=0.001$ are the resonance frequencies.}
    \label{fig7}
\end{figure*}

\begin{figure}[hbp]
\includegraphics[width=\columnwidth]{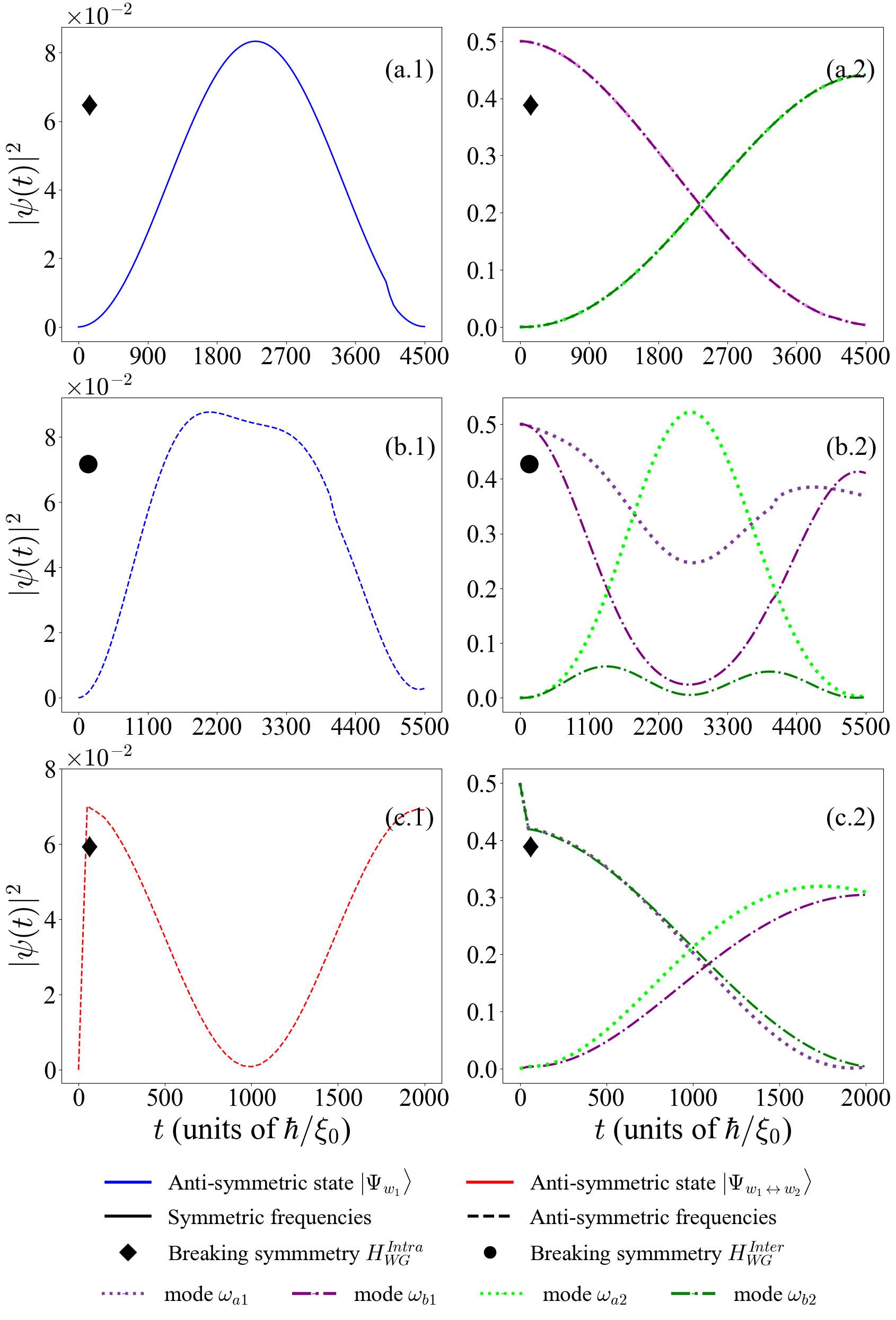}
      \caption{Probability density from the initial conditions for the anti-symmetric states ($|\Psi_{w_1}\rangle$ and $|\Psi_{w_{1}\leftrightarrow{w_2}}\rangle$) as a function of the time $t$ (Breaking symmetry with $H^{\text{Intra}}_{\text{WG}}$ and $H^{\text{Inter
      }}_{\text{WG}}$). The probability density for two cases: between the WGRs (left panels) and for the modes of the WGRs (right panels). The temporal evolution is shown up to $t=4500$ [(a.1) and (a.2)], up to $t=2000$ [(b.1) and (b.2)], and up to $t=2000$ [(c.1) and (c.2)] (all $t$ in units of $\hbar / \xi_{0}$). The parameters are set as $\xi_1=0.1$ and $\xi_0=1$ (both in units of $\xi_0$), with $\omega_{c}=0$. For symmetric frequencies, $\omega_{a1}=\omega_{b1}= \omega_{a2}=\omega_{b2}=0.001$, while for anti-symmetric frequencies, $\omega_{a1}=-\omega_{b1}= \omega_{a2}=-\omega_{b2}=0.001$ are resonance frequencies.}
    \label{fig8.a}
\end{figure}

We analyze the initial conditions $|\Psi_{w_1}\rangle$ and $|\Psi_{w_{1}\leftrightarrow{w_2}}\rangle$ states, where symmetry breaking occurs in the WGRs at $\Delta=0.00075$. For this analysis, we have selected parameters with $N=4000$ and $L=40$ to study these states in both their $H^{\text{Intra}}_{\text{WG}}$ and $H^{\text{Inter}}_{\text{WG}}$ configurations. When adjusting the state $|\Psi_{w_1}\rangle$ with breaking-symmetry $H^{\text{Intra}}_{\text{WG}}$ in Eq.\,(\ref{3.4.a}) as depicted in Fig.\,\ref{fig7}, we notice the formation of a trapped state between the WGRs, as shown in Fig.\,\ref{fig8.a}(a.1). This state exhibits characteristics similar to those observed in Fabry-Pérot interferometers and the interactions of BIC in Fabry-Pérot resonators \cite{Hsu2016,Shubin2023}. Moreover, the frequency modes of the WGRs are found to be aligned for each resonator, as indicated in Fig.\,\ref{fig8.a}(a.2). Similarly, it can be observed that the distribution of probability across the WGR modes is not entirely symmetrical, as shown in Fig.\,\ref{fig8.a}(a.2). This asymmetry is attributed to a portion of the probability being dispersed in the waveguide and the first WGR.

\begin{figure*}[htbp]
\includegraphics[width=\textwidth]{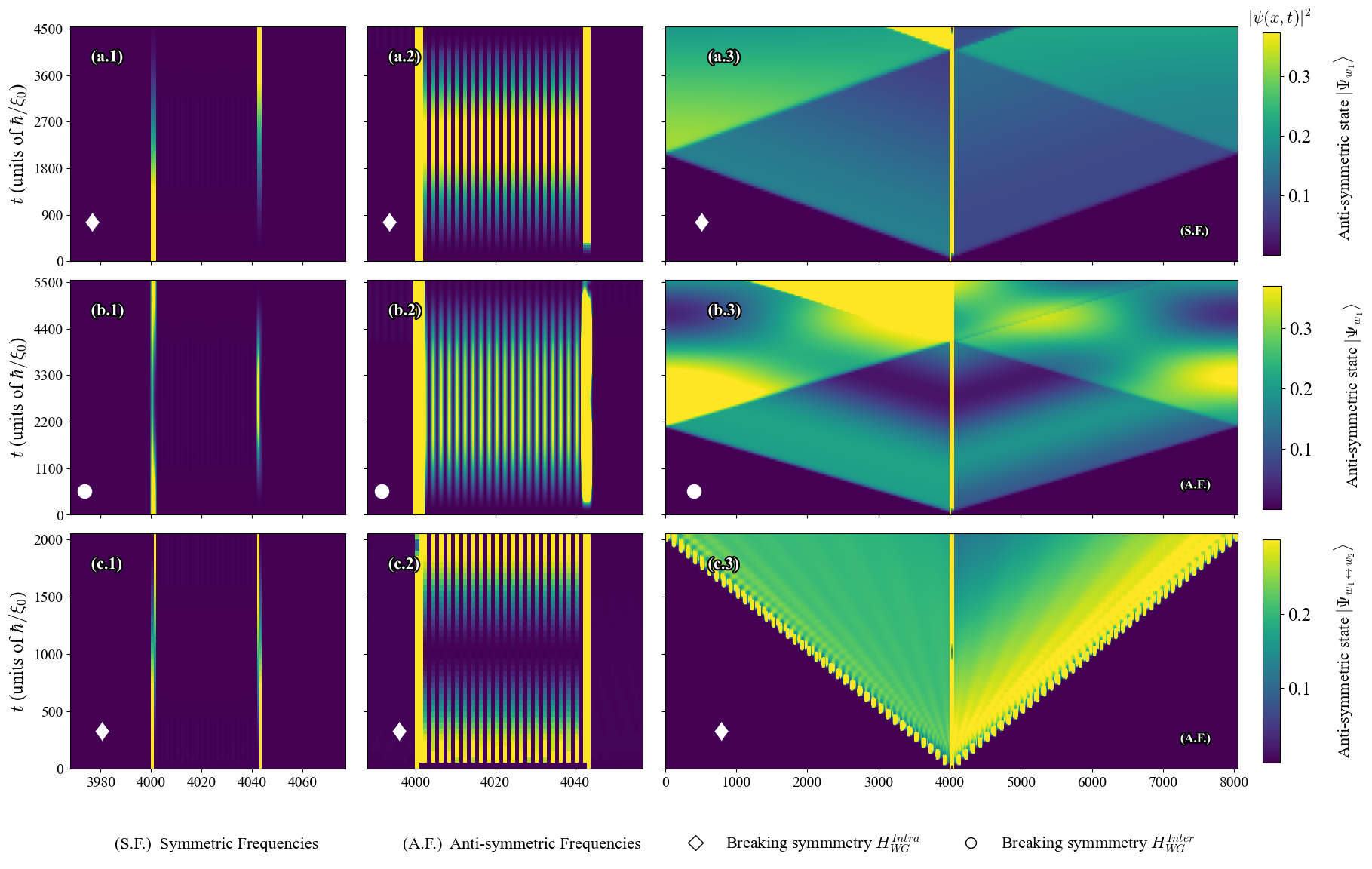}
      \caption{Dispersion properties from the initial conditions for the anti-symmetric states ($|\Psi_{w_1}\rangle$ and $|\Psi_{w_{1}\leftrightarrow{w_2}}\rangle$) as a function of the system modes. The figure shows the probability density of the wave packet as a function of position and time (Breaking symmetry with $H^{\text{Intra}}_{\text{WG}}$ and $H^{\text{Inter}}_{\text{WG}}$). We present the probability density $|\psi(x,t)|^2$ for three parts: first, within the WGRs; second, between the WGRs; and third, regarding the dispersion properties of anti-symmetric states within the WGRs. The temporal evolution is shown up to $t=4500$ [(a.1), (a.2), and (a.3)], up to $t=5500$ [(b.1), (b.2), and (b.3)], and up to $t=2000$ [(c.1), (c.2), and (c.3)] (all $t$ in units of $\hbar / \xi_{0}$).  The parameters are set as $\xi_1=0.1$ and $\xi_0=1$ (both in units of $\xi_0$), with $\omega_{c}=0$. For symmetric frequencies, $\omega_{a1}=\omega_{b1}= \omega_{a2}=\omega_{b2}=0.001$, while for anti-symmetric frequencies, $\omega_{a1}=-\omega_{b1}= \omega_{a2}=-\omega_{b2}=0.001$ are the resonance frequencies}
    \label{fig9.a}
\end{figure*}

In addition, Fig.\,\ref{fig9.a}(a.2) underlines the effective communication between the WGRs in our setup, with the most significant intensity of this interaction being observed around the median time, as seen in Fig.\,\ref{fig7}(a.3). Simultaneously, Fig.\,\ref{fig9.a}(a.1) reveals the cyclical process of filling and draining within the WGRs, observable at both the initial and final times as illustrated in Fig.\,\ref{fig7}. Complementing these observations, Fig.\,\ref{fig9.a}(a.3) displays a gradient of interaction over time, suggesting a gradual change in the system's state. This visual representation may correspond to the evolution of the resonance characteristics within the WGRs, indicating a sustained energy or information transfer process. The transition from darker to lighter hues denotes the system's passage from its nascent phase to a fully engaged state, depicting the dynamic behavior inherent to the WGR system.

Another scenario of interest involving the first set of WGRs emerges when breaking-symmetry with $H^{\text{Inter}}_{\text{WG}}$ in Eq.\,(\ref{3.4.b}), is applied, where the probability amplitudes and frequencies exhibit antisymmetry [refer to Fig.\,\ref{fig8.a}(b.2)]. In this evolution, we observe the occupancy number within the first WGR oscillating, as detailed in the evolution panel of Fig.\,\ref{fig7} [panels (b.1), (b.2), (b.3), and (b.4)]. Additionally, for the state trapped between the WGRs, shown in Fig.\,\ref{fig8.a}(b.1) and \ref{fig9.a}(b.2), its probability amplitude maintains a maximum value for a longer duration compared to its counterpart in the scenario with $H^{\text{Intra}}_{\text{WG}}$ in Eq. (\ref{3.4.a}), [see Appendix \ref{App.E},  Fig.\,\ref{Ap.fig2}(c) and \ref{Ap.fig3}(b)].


The introduction of antisymmetry in the frequencies once again results in a decoupling of the occupation modes within the WGRs [similarly in Appendix \ref{App.E} for anti-symmetric frequencies, see Fig.\,\ref{Ap.fig2}(d)]. Focusing on the mode occupations of the WGRs, as illustrated in Fig.\,\ref{fig8.a}(b.2), it becomes apparent that the mode occupation associated with the second WGR exhibits cyclic and symmetrical behavior. In other words, the occupation probabilities throughout the entire evolution period start and end at the same values. Conversely, for the first WGR [see Fig.\,\ref{fig8.a}(b.2)], we observe that the mode occupation does not exhibit complete symmetry, as a portion of this probability remains engaged in the interaction with the waveguide.


The observed phenomena indeed support the hypothesis that a initial condition antisymmetric state $|\Psi_{w_1}\rangle$ and corresponding frequencies, when combined with a symmetry-breaking interaction $H^{\text{Intra}}_{\text{WG}}$ as in Eq.\,(\ref{3.4.a}), can effectively facilitate information transfer via the WGRs. We observe a transfer fidelity in Eq.\,(\ref{3.5}) of the initial state in the first WGR to the second WGR of about $9/10$, as shown in Fig.\,\ref{fig7}(a.4). We have also checked that this value of the transfer fidelity is quite robust to changes in the distance between WGRs.

In contrast, for an anti-symmetric state $|\Psi_{w_1}\rangle$ and corresponding frequencies, combined with a symmetry-breaking interaction $H^{\text{Inter}}_{\text{WG}}$ as in Eq.\,(\ref{3.4.b}), the WGRs also effectively facilitate information transfer, acting as both emitter and receiver. However, this is the only scenario where the transfer fidelity is nearly zero, due to the probability amplitude remaining almost entirely within the modes of the first WGR. In this case, we obtain a fidelity of roughly $0.82$ for the state's probability amplitude. Supportive evidence shows the state's initial and final evolution and Fig.\,\ref{fig7}(b.1) and (b.4), which depicts the occupancy oscillations in the WGRs and the eventual restoration in the first WGR.

In our final scenario, we investigate the variation of the initial condition anti-symmetric state $|\Psi_{w_{1}\leftrightarrow w_2}\rangle$, while establishing antisymmetry in the frequencies $(\omega_{a1} = -\omega_{b1} = \omega_{a2} = -\omega_{b2})$ when breaking symmetry with $H^{\text{Inter}}_{\text{WG}}$ as described in Eq.\,(\ref{3.4.b}). The results, presented in Fig.\,\ref{fig7} [panels (c.1), (c.2), (c.3), and (c.4)], demonstrate that at the conclusion of the system's evolution, the transfer fidelity using the state $|\Psi_{w_{1}\leftrightarrow w_2}\rangle$ is about 3/5. Furthermore, the interaction between the WGRs exhibits behavior analogous to an anti-Fabry-Pérot effect, as shown in Fig.\,\ref{fig8.a}(c.1). This behavior contrasts with the Fabry-Pérot-like interactions observed in the previously evaluated WGR occupancies, referenced in Figs.\,\ref{fig8.a}(a.1) and \ref{fig8.a}(b.1).

This anti-Fabry-Pérot characteristic signifies a more intense communication between the WGRs towards the final time of the system's evolution, as shown in Fig.\,\ref{fig7}(c.4). Additionally, this behavior fosters a cyclic nature in the evolution of the state trapped between the WGRs [see Fig.\,\ref{fig8.a}(c.1)]. The anti-Fabry-Perot effect also ensures that the WGRs retain a significant portion of the system's amplitude throughout the system's evolution. This is while the remainder of the probability engages in the interaction between the WGRs and the waveguide, as observed in Fig.\,\ref{fig7}, indicating that the probability amplitude for the WGRs is not fully recovered. Additionally, this anti-Fabry-Pérot behavior has a characteristic signature in dispersion properties of a wave packet for the initial condition anti-symmetric state $|\Psi_{w_{1}\leftrightarrow{w_2}}\rangle$ as we can see in Fig.\,\ref{fig9.a}(c.3). Another crucial observation is that when antisymmetry is imposed on the WGR frequencies results in the decoupling of the WGR modes and diminishes the probability amplitudes, as illustrated in Fig.\,\ref{fig8.a}(c.2). This behavior is consistent with another initian condition anti-symmetrical state $|\Psi_{w_{1}}\rangle$, as depicted in Figs.\,\ref{fig8.a}(b.2) and \ref{Ap.fig2}(d) (see Appendix \ref{App.E}).

\section{Summary}
\label{sec:summary}
We studied single-photon transport through two WGRs coupled to a one-dimensional waveguide and discovered the emergence of BIC between the WGRs. Using Green's function formalism, we calculated the transmission spectra and the photonic DOS. The BIC is visualized as Dirac delta functions and Fano resonances in the DOS when the separation distance $L$ between the WGRs is even. Additionally, we found that the BICs appear in the photonic transmission as sharp resonances. By breaking the frequency symmetry, we obtained BICs for even $L$. Furthermore, we investigated the time evolution of an incident single-photon wave packet interacting with the WGRs. We analyzed the mechanism of partial storage of the wave packet in the BIC. 

The analysis of the DOS as a function of energy and site number between the ($L$) reveals distinct sharp peaks at even $L$ values, indicating the presence of BICs. These peaks, resulting from constructive interference, lead to highly localized states that do not radiate energy, underscoring the system's inherent symmetry. This symmetry fosters the formation of BICs, establishing a pattern of resonance at even $L$ intervals. The extremely high DOS values around the Fano resonance further emphasize the critical interplay between WGR pairs, which is essential for precise light manipulation in photonic device design. A deep understanding of these dynamics is essential for successfully designing and optimizing photonic systems incorporating WGRs. This knowledge enables precise control over the optical properties of the system, paving the way for advancements in light manipulation and the development of innovative photonic devices.

We also considered the initial conditions with a photon prepared in the anti-symmetric states $| \Psi_{w_1}\rangle$ and $|\Psi_{w_{1} \leftrightarrow w_2}\rangle$, breaking the symmetry with $H^{\text{Intra}}_{\text{WG}}$ and $H^{\text{Inter}}_{\text{WG}}$, and analyzed its time evolution. In both anti-symmetric states, a state emerges between the WGRs. The state $|\Psi_{w_1}\rangle$ exhibits behavior typical of a Fabry-Pérot interferometer, where the distance between the resonators or emitters is even. On the other hand, the state \(|\Psi_{w_{1} \leftrightarrow w_2}\rangle\) demonstrates the interaction between the WGRs that can be described as anti-Fabry-Pérot, contrasting with the Fabry-Pérot-like interactions observed in the previous state. This analysis underscores the intricate interplay between system components, paving the way for enhanced control over photonic system properties by strategically manipulating resonance and antiresonance behaviors.

For the initial condition  with a anti-symmetric state $|\Psi_{w_1}\rangle$, we observe two significant behaviors related to symmetric and anti-symmetric frequencies when breaking the symmetry with $H^{\text{Intra}}_{\text{WG}}$ and $H^{\text{Inter}}_{\text{WG}}$. By the end of the temporal evolution, the WGRs exhibit a trasnfer fidelity exceeding 4/5. For symmetric frequencies, the second WGR is predominantly occupied, while the first WGR is primarily occupied for anti-symmetric frequencies. Additionally, for anti-symmetric frequencies, the WGR modes decouple compared to the symmetric case. 
For the other intinial condition for the anti-symmetric state $|\Psi_{w_{1} \leftrightarrow w_2}\rangle$, both WGRs are occupied at the end of the evolution, with a trasnfer fidelity exceeding 3/5, for anti-symmetric frequencies. This analysis is essential for designing and creating photonic devices, such as optical filters, lasers, and sensors, where precise control over light propagation and resonance is of paramount importance \cite{Sadriev2019}. Because, we can use the system to send a single photon, with the two WGRs serving as receivers. For initial condition for the anti-symmetric state $|\Psi_{w_1}\rangle$, the system achieves a fidelity of 4/5 of the initial state under anti-symmetric frequencies. Another application is for initial condition for anti-symmetric state $|\Psi_{w_{1} \leftrightarrow w_2}\rangle$, where we can send entangled information into the two WGRs with a trasnfer fidelity of 3/5. 
These capabilities are particularly relevant for quantum information applications, such as quantum communication and quantum memory, where high trasnfer fidelity and the ability to control and manipulate quantum states are crucial. The system's ability to maintain high fidelity makes it suitable for secure information transfer, quantum entanglement distribution, and the implementation of quantum networks \cite{Chai2024}.

\acknowledgments
A.R.L. acknowledges financial support from a PIIC-UTFSM grant and ANID-Subdirecci\'{o}n de Capital Humano Doctorado Nacional Grant 2023-21230847. M.~A. acknowledges financial support from Postdoctoral FONDECYT Grant No. 3240443. J.P.R.-A is grateful for the financial support of FONDECYT Iniciaci\'on grant No. 11240637. P.A.O. acknowledges support from FONDECYT Grant No. 1180914 and 1201876. R.A.M. acknowledges financial support from
the Ministerio de Ciencia e Innovación of Spain (Spanish Ministry of Science, Innovation, and Universities) and
FEDER (ERDF: European Regional Development Fund) under Research Grant No. PID2022-136285NB-C31.

\appendix
\section{Green's function approach Transmission Amplitude: An In-Depth Analysis:}
\label{App.A}

This appendix demonstrates how, by using the Lippmann-Schwinger and Dyson equations, we calculate Green's function at sites $0$ and $L$ of the WGRs coupled to the waveguide to obtain the transmission amplitude.

As an initial assumption, we assume, 
\begin{equation}
\phi_{j}= e^{i k j} \quad \text{and} \quad \phi_{w_1}=\phi_{w_2}=0,
\label{A.1}
\end{equation}
where $\phi_{w_1}$ and $\phi_{w_2}$ represent the first and second WGRs, respectively also we have the WGRs disconnected.

The interaction between the waveguide and the resonator modes, as presented in Eq.\,(\ref{1.c}), can be effectively expressed as  
\begin{eqnarray}
H_{\text{I}}^{(\text{eff})}&=&\xi_{1}(\hat{c}_{0}^ {\dag}\hat{c}_{w_1}+\hat{c}_{0}\hat{c}_{w_1}^{\dag}) \nonumber 
+\xi_{1}(\hat{c}_ {L}^{\dag}\hat{c}_{w_2}+\hat{c}_{L}\hat{c}_{w_2}^{\dag}) .\\
\label{A.2}
\end{eqnarray}
The methodology is formally equivalent to substituting the WGRs with two discrete impurities. This procedural simplification is employed solely to facilitate the mathematical process and does not influence the final result. This assertion is substantiated by reintroducing the WGRs' corresponding Green's functions after the analysis. Note that the choice of a positive sign for $H_{\text{I}}^{(\text{eff})}$ is arbitrary. The equations presented here would be recovered should the reader opt for a negative sign. This is due to the parity of the even sign, which ensures that the choice of sign does not affect the final result.


In this way, we write the Lippmann-Schwinger equation as 
\begin{equation}
|\Psi\rangle= |\phi\rangle + G \, H_{\text{I}}^{(\text{eff})} \,|\phi\rangle \, ,
\label{A.4}
\end{equation}
where $|\Psi\rangle$ represents the perturbed state, $|\phi\rangle$ is the unperturbed state, $G$ is the Green's function associated with the unperturbed Hamiltonian, and $H_{\text{I}}^{(\text{eff})}$ denotes the interaction Hamiltonian. The term $G \, H_{\text{I}}^{(\text{eff})} \, |\phi\rangle$ accounts for the first-order correction to the state due to the interaction. This formulation allows us to iteratively solve for $|\Psi\rangle$ in the presence of the perturbation by expanding the series. 

Given our objective to ascertain the transmission amplitude for the WGRs system, we employ the bra  $\langle L+1 |$ as derived from the preceding equation: $\langle L+1 |\Psi\rangle = \langle L+1 |\phi\rangle + \langle L+1 |G \, H_{\text{I}}^{(\text{eff})} \,|\phi\rangle$. Now, with respect to the WGRs the terms that make sense from Eq.\,(\ref{A.1}) are $\langle 0 |\phi\rangle$ and $\langle L |\phi\rangle$, obtaining $\langle L+1 |\Psi\rangle = \langle L+1 |\phi\rangle + \langle L+1 |G \, | w_1 \rangle \langle w_1 | \, H_{I}^{(\text{eff})} \, | 0 \rangle \langle 0 | \, |\phi\rangle +\langle L+1 |G \, | w_2 \rangle \langle w_2 | \, H_{\text{I}}^{(\text{eff})} \, | L \rangle \langle N | \, |\phi\rangle$, finally 

\begin{equation}
\Psi_{L+1} = \phi_{L+1} + \xi_{1} \, G_{L+1,w_1} \,  \phi_{0} + \xi_{1} \, G_{L+1,w_2} \,  \phi_{L} \, .
\label{A.5}
\end{equation}
By determining $\Psi_{L+1}$ completely, we have to evaluate the amplitude of the corresponding photon, which is
\begin{equation}
\Psi_{j}= t e^{i k j} \, ,\quad \, j > L \, ,
\label{A.6}
\end{equation}
and in this way, obtain the transmission amplitude.

For the purpose of deriving $\Psi_{L+1}$, it is necessary to calculate the Green's functions for $G_{L+1,w_1}$ and $G_{L+1,w_2}$. This is accomplished by employing the Dyson equation,
\begin{equation}
G= g +g \,  H_{\text{I}}^{(\text{eff})} \, G .
\label{A.7}
\end{equation}

Illustrating a little the use of Dyson's equation to determine $G_{L+1,w_1}=\langle L+1 | G |w_1\rangle$, like this $G_{L+1,w_1}=\langle L+1 |( g +g \,  H_{\text{I}}^{(\text{eff})} \, G )|w_1\rangle= \langle L+1 |g \,  H_{\text{I}}^{(\text{eff})} \, G |w_1\rangle$ where the term associated with the matrix elements of the Green's function for the WGRs are only diagonal, since we have considered that they are not coupled to the waveguide [Eq.\,(\ref{A.1})]. We are left to consider the sites where the WGRs are located; therefore,
\begin{eqnarray}
G_{L+1,w_1} &&=  \langle L+1 |g \,| 0 \rangle \langle 0 |  H_{\text{I}}^{(\text{eff})} \, | w_1 \rangle \langle w_1 | G |w_1 \rangle \nonumber \\
&&+ \langle L+1 |g \,| L \rangle \langle L |  H_{\text{I}}^{(\text{eff})} \, | w_2 \rangle \langle w_2 | G |w_1 \rangle \\
&&= \xi_{1} ( \, g_{L+1,w_1} \, G_{w_1,w_1} +  g_{L+1,L} \, G_{w_2, w_1} )\, . \nonumber
\label{A.8}
\end{eqnarray}
Analogously, we obtain,
\begin{equation}
G_{L+1,w_2}= \xi_{1} \, ( g_{L+1,w_1} \, G_{w_1,w_2} +  g_{L+1,L} \, G_{w_2, w_2} ) \, .
\label{A.9}
\end{equation}
We have the Green's functions for the WGRs, which only have diagonal elements, and we define them as $g_{w_1,w_1}\equiv g_{w_1}$ and $g_{w_2,w_2}\equiv g_{w_2}$, written explicitly as
\begin{subequations}
\begin{eqnarray}
g_{w_1}=\frac{1}{\omega-\omega_{a1}+i\eta} +\frac{1}{\omega-\omega_{b1}+i\eta} \, , \label{A.10.a} \\
g_{w_2}=\frac{1}{\omega-\omega_{a2}+i\eta} +\frac{1}{\omega-\omega_{b2}+i\eta} \, ,
\label{A.10.b}
\end{eqnarray}  
\end{subequations}

for $\eta \rightarrow 0$. In addition,
\begin{equation}
g_{l,j}=\frac{e^{ik|l-j|}}{2i \, \xi_{0}\sin{(k)}} \, ,
\label{A.11}
\end{equation}
corresponding to the Green's function of the infinite chain \cite{Economou2006} or, in our case, the waveguide. 

We now identify the essential elements required to determine the Green's function $G_{L+1,w_1}$. We employ the Dyson equation to  derive $G_{w_1,w_1}$. For this purpose, the Dyson equation is utilized to calculate $G_{w_1,w_1}$, and this process is repeated iteratively until we derive a closed set of equations. This set enables us to solve the equations iteratively and thus determine $G_{L+1,w_1}$, resulting in:
\begin{subequations}
\begin{eqnarray}
G_{w_1,w_1}&&= g_{w_1} +\xi_{1} \, g_{w_1} \, G_{0,w_1} \, , \label{A.12.a} \\
G_{0,w_1}&&= \xi_{1} \, ( g_{0} \, G_{w_1,w_1} + g_{0,L} \, G_{w_2,w_1} ) \, , \label{A.12.b} \\
G_{w_2,w_1} &&= \xi_{1} \, g_{w_1} \, G_{L,w_1} \, ,\label{A.12.c} \\
G_{L,w_1}&&= \xi_{1}\, ( g_{L,0} \, G_{w_1,w_1} + g_{L} \, G_{w_2,w_1} ) \, .
\label{A.12.d}
\end{eqnarray}
\end{subequations}
From the previous expressions, we can see that we have defined the diagonal terms as $g_{0,0} \equiv g_{0}$ and $g_{L,L} \equiv g_{L}$.

Below we describe the resolution of the set of equations and some expressions necessary to obtain $G_{L+1,w_1}$. First, we introduce Eq.\,(\ref{A.12.d}) into Eq.\,(\ref{A.12.c}), obtaining
\begin{equation}
G_{w_{2},w_{1}}=\frac{g_{w_2} \, g_{L,0} \, \xi_{1}^{2}}{1-g_{w_1} \, g_{w_2} \xi_{1}^{2}} \, G_{w_{1},w_{1}} \, .
\label{A.13}
\end{equation}
Using this result in Eq.\,(\ref{A.12.b}), we obtain $G_{0,w_1}$ as a function of $G_{w_1,w_1}$, an expression which we simplify using
\begin{subequations}
\begin{eqnarray}
g_{0}&&=g_{L}=\frac{1}{2i \, \xi_{0}\sin{(k)}} \, , \label{A.14.a} \\
g_{0,L}&&=g_{L,0}= g_{0}^{2} \, e^{2ik|L|} \, . \label{A.14.b}
\end{eqnarray}
\end{subequations}

Now, with $G_{0, w_1}$ depending on $G_{w_1,w_1}$, we can introduce this result in Eq.\,(\ref{A.10.a}) and obtain
\begin{equation}
G_{w_1,w_1}= \frac{g_{w_1}\, (1- g_{0} \, g_{w_2} \, \xi_{1}^{2} )}{1-(g_{w_1}+g_{w_2})\Sigma_{0} +g_{w_1}g_{w_2}\Sigma_{0}^{2}(1-e^{2ik|L|}) } \, ,
\label{A.15}
\end{equation}
where $\Sigma_{0}=\xi_{1}^{2}g_{0}=\xi_{1}^{2}/2i\xi_{0}\sin(k)$ is the self-energy.

By combining the results obtained in Eqs.\,(\ref{A.13}) and (\ref{A.15}) with Eq.\,(\ref{A.8}), using
\begin{subequations}
\begin{eqnarray}
g_{L+1,0} \, g_{0}&&= g_{0}^{2} \, e^{ik|L+1|} \, , \label{A.16.a} \\
g_{L+1,L} \, g_{L,0}&&=g_{L,0}= g_{0}^{2} \, e^{2ik|L+1|} \, , \label{A.16.b}
\end{eqnarray}
\end{subequations} 
we obtain, 
\begin{equation}
G_{L+1,w_1}=\frac{g_{L+1,0}\, g_{w_1} \, \xi_{1}}{1-(g_{w_1}+g_{w_2})\Sigma_{0} +g_{w_1}g_{w_2}\Sigma_{0}^{2}(1-e^{2ik|L|})} \, .
\label{A.17}
\end{equation}

We must still determine $G_{L+1,w_2}$. To achieve this, we will employ a methodology analogous to the one used to derive 
$G_{L+1,w_1}$. Thus, the equations forming a comprehensive and closed set necessary for calculating $G_{L+1,w_2}$  are:
\begin{subequations}
\begin{eqnarray}
G_{w_2,w_2}&&= g_{w_2} +\xi_{1} \, g_{w_2} \, G_{L,w_2} \, \label{A18.a} \\
G_{L,w_2}&&= \xi_{1} \, ( g_{L,0} \, G_{W_1,w_2} +g_{L} \, G_{W_2,w_2}) \, \label{A18.b} \\
G_{w_1,w_2}&&= \xi_{1} \, g_{w_1} \, G_{0,w_2}  \, \label{A18.c} \\
G_{0,w_2}&&= \xi_{1} \, ( g_{0} \, G_{W_1,w_2} +g_{0,L} \, G_{W_2,w_2}) \, .
\label{A18.d} 
\end{eqnarray}
\end{subequations}
By solving this set of equations and using the expressions
\begin{subequations}
\begin{eqnarray}
g_{L+1,0} \, g_{0,L}&&= g_{0}^{2} \, e^{ik|L+1|} \, e^{ik|L|}  \, , \label{A.19.a} \\
g_{L+1,L} \, g_{0}&&=g_{L,0}= g_{0}^{2} \, e^{ik} \, , \label{A.19.b}
\end{eqnarray}
\end{subequations} 
thereby obtaining the expression
\begin{small}
\begin{equation}
G_{L+1,w_2}=\frac{g_{L+1,L}\, g_{w_2} \, \xi_{1} -g_{w_1}\, g_{0}^{2}\, \xi_{1}^{3} \, (e^{ik|1|}-e^{ik|L+1|} \, e^{ik|L|})}{1-(g_{w_1}+g_{w_2})\Sigma_{0} +g_{w_1}g_{w_2}\Sigma_{0}^{2}(1-e^{2ik|L|})} \, .
\label{A.20}
\end{equation}
\end{small}
By introducing the expressions obtained in Eqs.\,(\ref{A.17}) and (\ref{A.20}) into Eq.\,(\ref{A.5}), in addition to making use of  Eq.\,(\ref{A.6}) and $\phi _{j}=e^{ikj}$, we can simplify the expression using,
\begin{subequations}
\begin{eqnarray}
g_{L+1,0} \, \phi_{0}&&= g_{0} \, e^{ik(L+1)}  \, , \label{A.21.a} \\
g_{L+1,L} \, \phi_{L}&&= g_{0} \, e^{ik(L+1)} \, ,
\label{A.21.b} 
\end{eqnarray}
\end{subequations} 
obtaining the transmission amplitude,
\begin{equation}
t=\frac{1}{1-(g_{w_1}+g_{w_2})\Sigma_{0} +g_{w_1}g_{w_2}\Sigma_{0}^{2}(1-e^{2ikL})} \, ,
\label{A.22}
\end{equation}
where $\Sigma_0$, $g_{w_1}$ and $g_{w_2}$ are the quantities defined above.

\section{Green's function approach for WGRs: An In-Depth Analysis:}
\label{App.B}
In this appendix, we determine Green's function for the WGRs located at $0$ and $L$ coupled to the waveguide (see Fig. \ref{fig1}).

Starting with the interaction Hamiltonian as depicted in Eq.\,(\ref{1.c}), and then performing a trace operation on the Dyson equation for the modes of the first WGR, $G_{l,l}=\langle l |( g +g \,  H_{\text{I}} \, G ) | l \rangle$, we obtain:
\begin{equation}
G_{l,l}= g_{l} -\xi \, g_{l} \, G_{0,l} \, ,
\label{B.1}
\end{equation}
where $l=\omega_{a1}, \omega_{b1}$  represent the modes of the first WGR and $g_{l,l} \equiv g_{l}$. By iteratively applying the Dyson equation, we derive the closed sets of equations necessary to determine $G_{l,l}$, These sets are as follows:
\begin{subequations}
\begin{eqnarray}
G_{0,l}&&=  -\xi\, (g_{0} \, G_{l,l} + g_{0,L} \sum_{m} G_{m,l}) \, , \label{B.2.a} \\
G_{m,l}&&=  -\xi\, g_{m} \, G_{L,l}  \, , \label{B.2.b} \\
G_{0,l}&&=  -\xi\, (g_{L,0} \, G_{l,l} + g_{L} \sum_{m} G_{m,l}) \, , 
\label{B.2.c}
\end{eqnarray}
\end{subequations}
where $m=\omega_{a2}, \omega_{b2}$  represent the modes of the second WGR and $g_{m,m} \equiv g_{m}$. With this set of equations, we calculate $G_{l,l}$, where the final expression is simplified by utilizing $g_{0,L}=g_{L,0}= g_{0}^{2} \, e^{2ikL}$. Consequently, we obtain,
\begin{equation}
G_{l,l}= \frac{g_{l}\, (1-\sum\limits_{m}  g_{m} \, g_{0} \, \xi_{1}^{2})}{1-(g_{l}+\sum\limits_{m} \, g_{m})\Sigma_{0} +g_{l}\, \sum\limits_{m}  g_{m} \Sigma_{0}^{2}(1-e^{2ikL})} \, .
\label{B.3}
\end{equation}
Thus, it follows that \( \sum\limits_{m} g_{m} = \sum\limits_{m} 1/(\omega - m + i \eta) \) represents the summation over the modes \(m=\omega_{a2}, \omega_{b2}\) of the second WGR. Consequently, equating \(\sum\limits_{m} g_{m}\) to \(g_{w_2}\) provides us with the framework to derive the Green's function for the first WGR, as outlined below:
\begin{equation}
G_{l,l}= \frac{g_{l}\, (1-  g_{w_2} \, g_{0} \, \xi_{1}^{2})}{1-(g_{l}+g_{w_2})\Sigma_{0} +g_{l}\,   g_{w_2} \Sigma_{0}^{2}(1-e^{2ikL})} \, ,
\label{B.4}
\end{equation}

A similar procedure can be employed to derive the Green's function for the second WGR. Alternatively, it can be obtained by substituting $l$ to $m$ in the Eq.\,(\ref{B.3}) and applying the relationship $\sum\limits_{l}  g_{l}=\sum\limits_{l} 1/(\omega -l +i \eta)= g_{w_1} $ and $l=\omega_{a1}, \omega_{b1}$. Implementing this adjustment provides us with the Green's function for the second WGR as follows:
\begin{equation}
G_{m,m}= \frac{g_{m}\, (1-  g_{w_1} \, g_{0} \, \xi_{1}^{2})}{1-(g_{m}+g_{w_1})\Sigma_{0} +g_{m}\,   g_{w_1} \Sigma_{0}^{2}(1-e^{2ikL})} \, .
\label{B.5}
\end{equation}

\section{Green's function approach between WGRs: An In-Depth Analysis:}
\label{App.C}
In this appendix, utilizing the Dyson equation, we calculate the Green's function between the sites $0$ and $L$, where the WGRs are coupled to the waveguide.

Now, we calculate the diagonal elements over the number of sites in the waveguide, denoted by $j$. This is achieved by using the Dyson equation $G_{j,j}=\langle j |( g +g \, H_{\text{I}}^{(\text{eff})} \, G ) | j \rangle$, employing the Hamiltonian detailed in Eq.\,(\ref{A.2}),
\begin{equation}
G_{j,j}=g_{j} +\xi_{1}\, (g_{j,0} \, G_{w_1,j} + g_{j,L} \, G_{w_2,j} ) \, ,
\label{C.1}
\end{equation}
where $g_{j,j} \equiv g_{j}$. By iteratively applying the Dyson equation, we derive a comprehensive set of closed equations, enabling us to determine $G_{j,j}$. These equations are as follows:
\begin{subequations}
\begin{eqnarray}
G_{w_1,j}&&=\xi_{1} \, g_{w_1} \, G_{0,j} \, , \label{C.2.a} \\
G_{w_2,j}&&=\xi_{1} \, g_{w_2} \, G_{L,j} \, , \label{C.2.b} \\
G_{0,j}&&=g_{0,j} +\xi_{1} \, ( g_{0} \, G_{w_1,j} + g_{0,L} \, G_{w_2,j} )\, , \label{C.2.c} \\
G_{L,j}&&=g_{L,j} +\xi_{1} \, ( g_{L,0} \, G_{w_1,j} + g_{L} \, G_{w_2,j} )\, . \label{C.2.c} 
\end{eqnarray}
\end{subequations}
By consistently employing this set of equations, we can calculate 
$G_{j,j}$, for which the following simplifications apply:
\begin{subequations}
\begin{eqnarray}
g_{j,0}&&=g_{0,j}= g_{0} \, e^{ikj} \, \label{C.3.a} \\
g_{L,0}&&=g_{0,L}= g_{0} \, e^{ikL} \, \label{C.3.b} \\
g_{L,j}&&=g_{j,L}= g_{0} \, e^{ik(L-j)} \, \label{C.3.c} \\
\end{eqnarray}
\end{subequations}
Given the range of sites $0 \leq j \leq L$, our objective is to calculate the Green's function connecting the WGRs. This function  is formulated as:
\begin{equation}
G_{j,j}= \frac{ g_{0}  \left( 1 -F_{j} \right) }{1-(g_{w_1}+g_{w_2})\Sigma_{0} +g_{w_1}\,   g_{w_2} \Sigma_{0}^{2}(1-e^{2ikL})} \, ,
\label{C.4}
\end{equation}
where $F_{j}$ is defined as
\begin{eqnarray}
F_{j} & \equiv g_{w_1} \, \Sigma_{0} \, (1-e^{2ikj}) +g_{w_2} \, \Sigma_{0} \, (1-e^{2ik(L-j)}) \nonumber \\
&- g_{w_1} \, g_{w_2} \,  \Sigma_{0}^{2} (1+e^{2ikL}) (1+e^{-2ikj}) \, .
\label{C.5}
\end{eqnarray}
This Eq. (\ref{C.4}) represents the Green's function for WGRs situated across sites numbered from $0$ to $L$, capturing the complex interactions between these resonators.

\section{ Transmission Amplitude with wave function: An In-Depth Analysis}
\label{App.D}
In this appendix, we present an alternative approach to derive the transmission expression, achieving the same results as those obtained using Green's functions. 

Analyzing single excitation propagation through the system involves examining the energy spectrum of the Hamiltonian $H$. This spectrum is determined by considering the eigenstate of the single excitation  as 
\begin{eqnarray}
|\Psi_E\rangle&=&\sum_{j}u^{k}_{j}\,\hat{c}_{j}^{\dag}|0\rangle+u_{a1}\,\hat{a}_{1}^{\dag}|0\rangle+u_{b1}\,\hat{b}_{1}^{\dag}|0\rangle \nonumber\\
&&+\,u_{a2}\,\hat{a}_{2}^{\dag}|0\rangle+u_{b2}\,\hat{b}_{2}^{\dag}|0\rangle\, .
\label{D.1}
\end{eqnarray}

We obtain the following stationary equations for the amplitudes from the eigenvalue equation $H|\Psi_E\rangle=E|\Psi_E\rangle$, 

\begin{subequations}
\begin{align}
&(\omega_{c}-\omega)\,u^{k}_{j} = \xi_{0}(u^{k}_{j+1}+u^{k}_{j-1})+\xi_{1}(u_{a1}+u_{b1})\,\delta_{j,0}\nonumber\\
&+\,\xi_{1}(u_{a2}+u_{b2})\,\delta_{j,L} \,,\\
&(\omega_{a1}-\omega)\,u_{a1}= \xi_{1}\,u^{k}_{0}\,, \\
&(\omega_{b1}-\omega)\,u_{b1}= \xi_{1}\,u^{k}_{0}\,,\\
&(\omega_{a2}-\omega)\,u_{a2}= \xi_{1}\,u^{k}_{L}\,, \\
&(\omega_{b2}-\omega)\,u_{b2}= \xi_{1}\,u^{k}_{L}\,.
\end{align}
\label{D.2}
\end{subequations}
Reducing the Eqs. (\ref{D.2}), we write
\begin{equation}
\left[\omega_{c}-\omega-V(j)\right]\,u^{k}_{j}=\xi_{0}(u^{k}_{j+1}+u^{k}_{j-1})\,,
\label{D.3}
\end{equation}
where $V(j)$ is the effective potential, defined as
\begin{eqnarray}
V(j)&=&\xi_{1}^{2}\left(\frac{1}{\omega_{a1}-\omega}+\frac{1}{\omega_{b1}-\omega}\right)\,\delta_{j,0}\nonumber \\
&+&\,\xi_{1}^{2}\left(\frac{1}{\omega_{a2}-\omega}+\frac{1}{\omega_{b2}-\omega}\right)\,\delta_{j,L}\, , \nonumber \\
&=& \xi_{1}^{2} ( v_0 \, \delta_{j,0} + v_L \, \delta_{j,L}) \, .
\label{D.4}
\end{eqnarray}

Under the standard scattering boundary conditions, a plane wave incident from $-\infty$ in the WGRs, the photon amplitude in the system can be written as
\begin{align}
u^{k}_{j}&=\ \begin{cases}
e^{ikj}+re^{-ikj}\ ,\qquad & j<0\ ,\\
Ae^{ikj}+Be^{-ikj}\ ,\qquad & 0<j<L\ ,\\
te^{ikj}\ ,\qquad & j>L\,,
\end{cases}
\label{D.5}
\end{align}
where $r$ and $t$ correspond to the reflection and transmission amplitudes, respectively, and $A$ and $B$ denote the backward and forward amplitudes in the central zone of the device. We proceed with the assumption of an unitary incident amplitude. The expression for the transmission amplitude is then given as:
\begin{equation}
t=-\frac{4 \xi_{0}^{2}\sin^{2}(k)}{ \left(2i\xi_{0} \sin(k)+v_{0}\right)\left(2i\xi_{0} \sin(k)+v_{L}\right)-e^{2ikL}\,v_{L} v_{0}}\,.
\label{D.6}
\end{equation}

The transmission probability, defined as $T =t \cdot t^{*} =|t|^{2}$, is given by the following equation:
\begin{equation}
T=\frac{4 \xi_{0}^{4}\sin^{4}(k)}{T^{2}_1(k) + T^{2}_2(k)} \, ,
\label{D.7}
\end{equation}
where $T_1(k) \equiv v_{0} v_{L}\sin^{2}(kL)-2\xi_{0}^{2}\sin^{2}(k)$ and $T_2(k) \equiv \xi_{0}(v_{0}+v_{L})\sin(k)-v_{0}v_{L}\sin(kL)\cos(kL)$. The Eq.\, (\ref{D.7}) analyze the transmission in devices, emphasizing the interplay between wave propagation, device architecture, and key factors: wave vector \(k\), effective potentials \(v_{0}\) and \(v_{L}\), and intrinsic properties \(\xi_{0}\). They are vital for assessing transmission efficiency and resonance in photonic systems, revealing the fundamental principles driving system performance.

\section{ Evolution for anti-symmetric state $|\Psi_{w_1}\rangle$ with anti-symmetric frequencies: An In-Depth Analysis}
\label{App.E}
In the final appendix, we analyze the temporal evolution of the initial condition for the anti-symmetric state \(|\Psi_{w_1}\rangle\) when the frequencies are anti-symmetric, mirroring its behavior under symmetric frequency conditions.
\begin{figure}[H]
\includegraphics[width=\columnwidth]{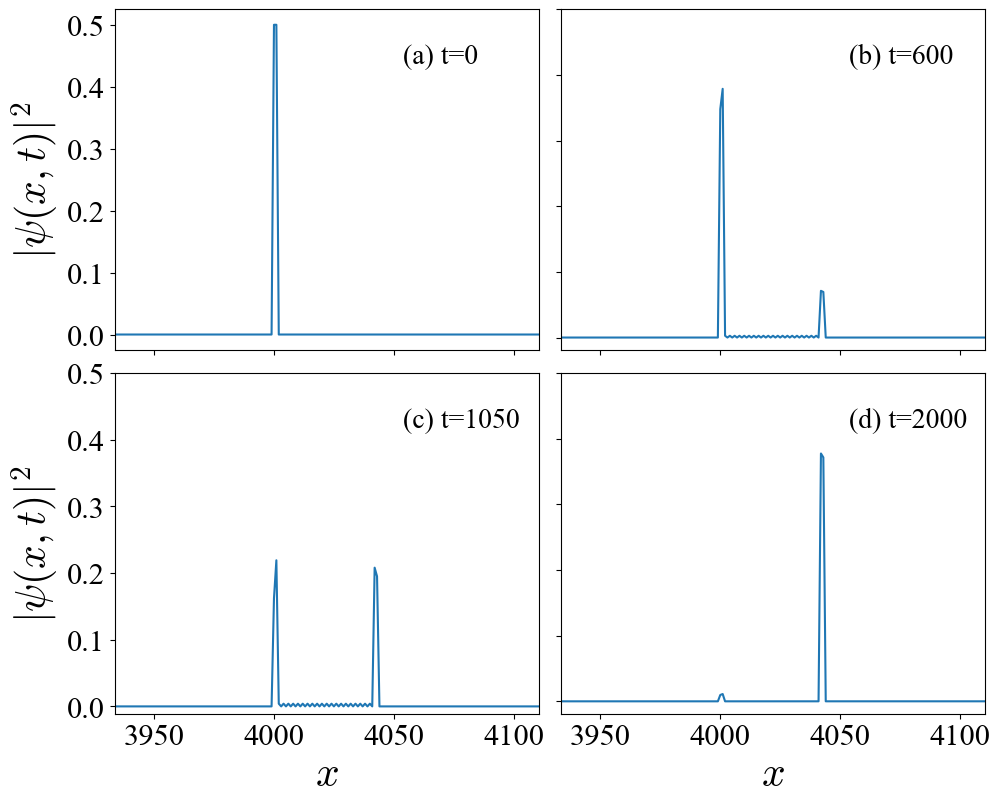}
      \caption{Probability density from the initial condition for the anti-symmetric state $|\Psi_{w_1}\rangle$ as a function of the system modes (Breaking symmetry with $H^{\text{Intra}}_{\text{WG}}$). The temporal evolution is shown for (a) $t=0$, (b) $t=600$, (c) $t=1050$ and (d) $t=2000$ (units of $\hbar / \xi_{0}$). The parameters are chosen as $\xi_1=0.1$ and $\xi_0=1$ (all in units of $\xi_0$), $\omega_{c}=0$, $\omega_{a1}=-\omega_{b1}= \omega_{a2}=-\omega_{b2}=0.001$ are resonance frequencies.}
    \label{Ap.fig1}
\end{figure}

\begin{figure}[H]
\includegraphics[width=\columnwidth]{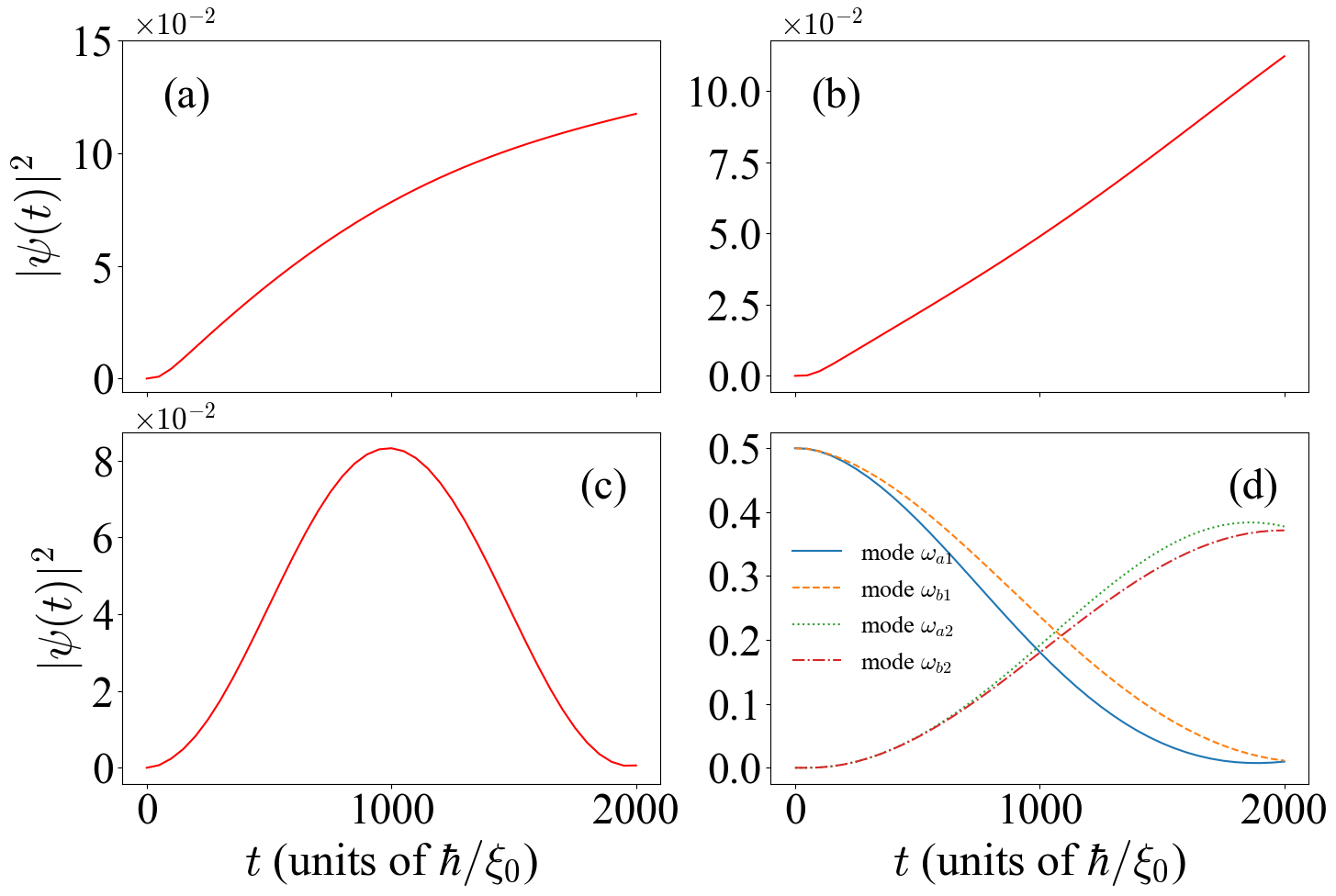}
      \caption{Probability density from the initial condition for the anti-symmetric state $|\Psi_{w_1}\rangle$ as a function of the time $t$ (Breaking symmetry with $H^{\text{Intra}}_{\text{WG}}$ ). The temporal evolution is shown up to $t=2000$ (units of $\hbar / \xi_{0}$). We present the probability density for: (a) the left side of the WGRs; (b) the right side of the WGRs; (c) between the WGRs, and (d) the modes of the WGRs. The parameters are chosen as $\xi_1=0.1$ and $\xi_0=1$ (all in units of $\xi_0$), $\omega_{c}=0$, $\omega_{a1}=-\omega_{b1}= \omega_{a2}=-\omega_{b2}=0.001$ are the resonance frequencies.}
    \label{Ap.fig2}
\end{figure}

\begin{figure}[H]
\includegraphics[width=\columnwidth]{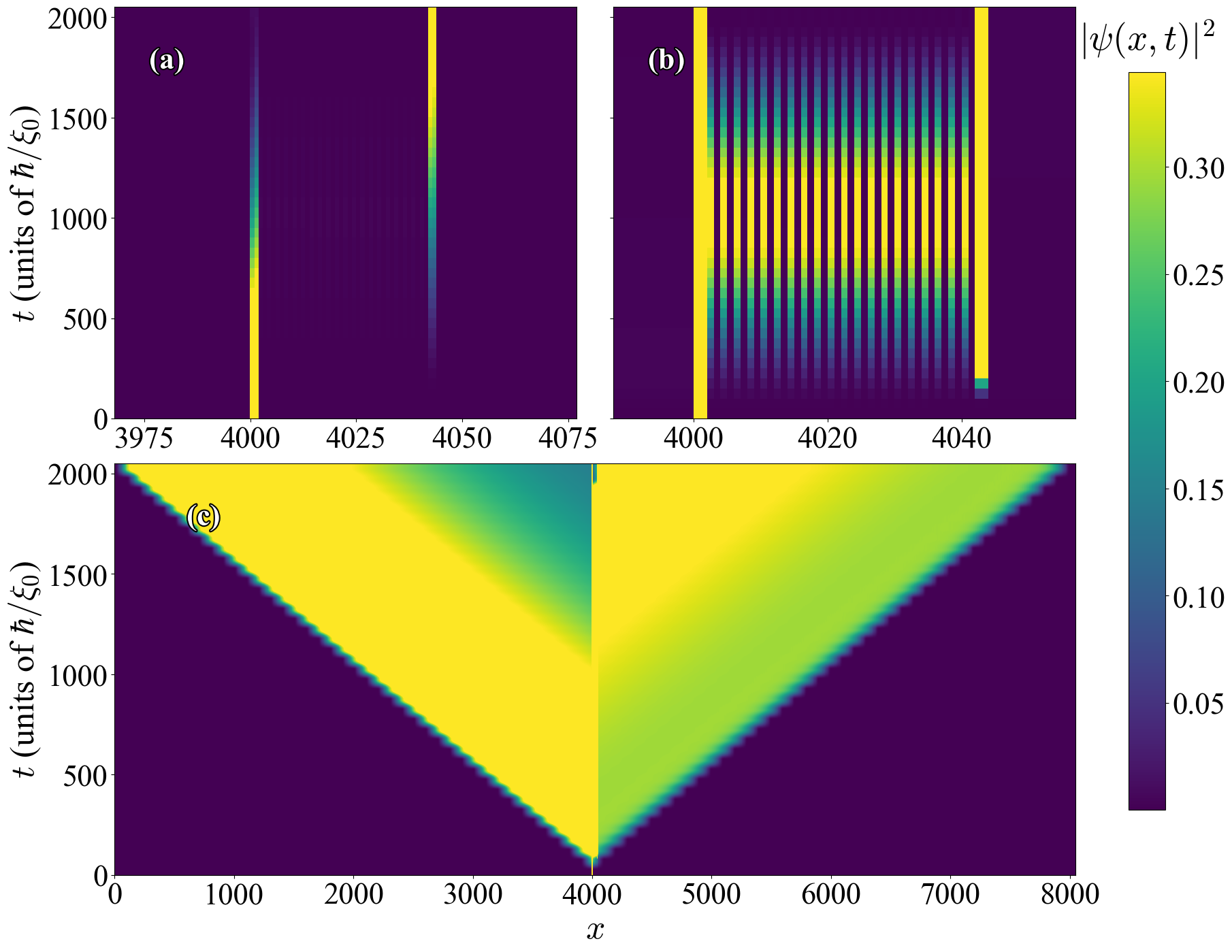}
       \caption{Dispersion properties from the initial condition for the anti-symmetric states $|\Psi_{w_1}\rangle$ as a function of the system modes. The figure shows the probability density of the wave packet as a function of position and time (Breaking symmetry with $H^{\text{Intra}}_{\text{WG}}$). The temporal evolution is shown up to $t=2000$ (units of $\hbar / \xi_{0}$). We present the probability density for: (a) the WGRs; (b) between the WGRs; and (c) dispersion properties of a wave packet for anti-symmetric state $|\Psi_{w_1}\rangle$ of the WGRs. The parameters are chosen as $\xi_1=0.1$ and $\xi_0=1$ (all in units of $\xi_0$), $\omega_{c}=0$, $\omega_{a1}=-\omega_{b1}= \omega_{a2}=-\omega_{b2}=0.001$ are the resonance frequencies.}
    \label{Ap.fig3}
\end{figure}

On the other hand, when we apply  anti-symmetric state $|\Psi_{w_1}\rangle$ to the frequencies of the WGRs, such that $\omega _{a1} = -\omega _{b1} = \omega _{a2} = -\omega _{b2}$, and introduce the same breaking-symmmetry with $H^{\text{Intra}}_{\text{WG}}$, the results depicted in Figs.\,\ref{Ap.fig1}, \ref{Ap.fig2}, and \ref{Ap.fig3} appear similar to those of their symmetrical counterparts in Figs.\,\ref{fig7}, \ref{fig8.a}, and \ref{fig9.a}. However, integrating antisymmetry into the WGR frequencies initiates a decoupling in the mode occupation probabilities of the WGRs, as illustrated in Fig.\,\ref{Ap.fig2}(d). In parallel, the probability density decreases compared to the scenario with symmetrical WGR frequencies, as evidenced in Fig.\,\ref{fig7}(a.4).

\twocolumngrid

\bibliographystyle{apsrev4-1}
\bibliography{references}

\end{document}